%% file: lrg_final.tex
\documentclass[preprint,aps,preprintnumbers]{emulateapj}

\usepackage{natbib}
\usepackage{graphicx}
\usepackage{dcolumn}
\usepackage{bm}
\usepackage{epsfig}
\usepackage{amsmath}
\usepackage{color}
\usepackage{longtable}
\usepackage[OT2,T1]{fontenc}

\newcommand{\Nb}{{\it N}-body}
\newcommand{\LRGs}{CMASS1}
\newcommand{\LRGe}{CMASS2}
\newcommand{\LRGn}{CMASS3}
\newcommand{\LRGt}{CMASS4}

\newcommand{\DArs}{D_A(z)/r_s}
\newcommand{\afit}[3]{#1^{+#2}_{-#3}}

\newcommand{\be}{\begin{equation}}
\newcommand{\ee}{\end{equation}}
\newcommand{\bea}{\begin{eqnarray}}
\newcommand{\eea}{\end{eqnarray}}

\def\thetaB{\mbox{\boldmath$\hat\theta$}}
\newcommand\gsim{\mathrel{\rlap{\lower4pt\hbox{\hskip1pt$\sim$}}
        \raise1pt\hbox{$>$}}}
\newcommand\lsim{\mathrel{\rlap{\lower4pt\hbox{\hskip1pt$\sim$}}
        \raise1pt\hbox{$<$}}}
\def\thetaB{\mbox{\boldmath$\hat\theta$}}

\begin{document}
\topmargin=24pt
\textheight=8.7in

\title{Clustering of Sloan Digital Sky Survey III Photometric Luminous
  Galaxies: The Measurement, Systematics and Cosmological Implications}

\author{
Shirley Ho\altaffilmark{1,2,3},
Antonio Cuesta\altaffilmark{4},
Hee-Jong Seo\altaffilmark{5},
Roland de Putter\altaffilmark{6,7},
Ashley J. Ross\altaffilmark{8},
Martin White\altaffilmark{1,9,10},
Nikhil Padmanabhan\altaffilmark{4},
Shun Saito\altaffilmark{10},
David J.\ Schlegel\altaffilmark{1},
Eddie Schlafly\altaffilmark{11},
Uros Seljak\altaffilmark{1,9,10,12,13},
Carlos Hern\'andez-Monteagudo\altaffilmark{14,26},
Ariel G. S\'anchez\altaffilmark{15},
Will J. Percival\altaffilmark{8}
Michael Blanton\altaffilmark{16},
Ramin Skibba\altaffilmark{17}
Don Schneider \altaffilmark{27,28}
Beth Reid\altaffilmark{1,29}
Olga Mena\altaffilmark{7},
Matteo Viel\altaffilmark{19,20}
Daniel J. Eisenstein \altaffilmark{11}
Francisco ~Prada\altaffilmark{21}
Benjamin Weaver\altaffilmark{16}
Neta Bahcall\altaffilmark{22}
Dimitry Bizyaev \altaffilmark{23}
Howard Brewinton \altaffilmark{23}
Jon Brinkman\altaffilmark{23}
Luiz Nicolaci da Costa\altaffilmark{24,25}
John R. Gott \altaffilmark{22}
Elena Malanushenko \altaffilmark{23}
Viktor Malanushenko \altaffilmark{23}
Bob Nichol \altaffilmark{8}
Daniel Oravetz \altaffilmark{23}
Kaike Pan \altaffilmark{23}
Nathalie Palanque-Delabrouille \altaffilmark{30}
Nicholas P Ross  \altaffilmark{1}
Audrey Simmons \altaffilmark{23}
Fernando de Simoni\altaffilmark{24,25,31}
Stephanie Snedden \altaffilmark{23}
Christophe Yeche \altaffilmark{30}
}
\altaffiltext{1}{Lawrence Berkeley National Laboratory, 1 Cyclotron Rd, MS 50R-5045, Berkeley, CA 94720, USA}
\altaffiltext{2}{Carnegie Mellon University, Physics Department, 5000 Forbes Ave, Pittsburgh, PA 15213, USA}
\altaffiltext{3}{cwho@lbl.gov}
\altaffiltext{4}{Yale Center for Astronomy and Astrophysics, Yale University, New Haven, CT 06511, USA}
\altaffiltext{5}{Berkeley Center for Cosmological Physics,LBL and Department of Physics, University of California, Berkeley, CA 94720, USA}
\altaffiltext{6}{ICC, University of Barcelona (IEEC-UB), Marti i Franques 1,Barcelona 08028, Spain}
\altaffiltext{7}{Instituto de Fisica Corpuscular, Universidad de Valencia-CSIC, Spain}
\altaffiltext{8}{Institute of Cosmology \& Gravitation, Dennis Sciama Building, University of Portsmouth, Portsmouth PO1 3FX, UK}
\altaffiltext{9}{Department of Physics,University of California Berkeley, Berkeley, CA}
\altaffiltext{10}{Department of Astronomy, University of California Berkeley, CA}
\altaffiltext{11}{Department of Astronomy, Harvard University, 60 Garden St. MS 20, Cambridge MA 02138}
\altaffiltext{12}{Institute for Theoretical Physics, University of Zurich,Winterthurerstrasse 190, CH-8057 Zurich, Switzerland}
\altaffiltext{13}{Ewha University, Seoul 120-750, Korea}
\altaffiltext{14}{Centro de Estudios de F\' \i sica del Cosmos de Arag\'on (CEFCA), Plaza de San Juan 1, planta 2, E-44001, Teruel, Spain}
\altaffiltext{15}{Max-Planck-Institut f\"ur Extraterrestrische Physik, Giessenbachstrasse 1,85748 Garching, Germany}
\altaffiltext{16}{Center for Cosmology and Particle Physics,Department of Physics,New York University,4 Washington Place, New York, NJ 10003, USA}
\altaffiltext{17}{Steward Observatory, University of Arizona, 933 N. Cherry Avenue, Tucson, AZ 85721, USA}
\altaffiltext{18}{Instituto de Fisica Corpuscular (IFIC), Universidad de Valencia-CSIC, Spain}
\altaffiltext{19}{INAF - Osservatorio Astronomico di Trieste, Via G.B. Tiepolo 11, I-34131 Trieste, Italy}
\altaffiltext{20}{INFN/National Institute for Nuclear Physics, Via Valerio 2, I-34127 Trieste, Italy}
\altaffiltext{20}{Instituto de Fisica Corpuscular (IFIC), Universidad de Valencia-CSIC, Spain}
\altaffiltext{21}{Instituto de Astrof\'{i}sica de Andaluc\'{i}a (CSIC), E-18080 Granada, Spain}
\altaffiltext{22}{Astrophysical Science, Princeton University, Princeton, NJ08544}
\altaffiltext{23}{Apache Point Observatory, P. O. Box 59, Sunspot, NM88349-0059}
\altaffiltext{24}{Laboratório Interinstitucional de e-Astronomia - LIneA, Rua Gal. José Cristino 77, Rio de Janeiro, RJ  20921-400, Brazil}
\altaffiltext{25}{Observatório Nacional, Rua Gal. José Cristino 77, Rio de Janeiro, RJ  20921-400, Brazil}
\altaffiltext{26}{Max-Planck-Institut f\"ur Astrophysik, Karl-Schwarzschild Str. 1,D-85748 Garching, Germany}
\altaffiltext{27}{Department of Astronomy and Astrophysics, The Pennsylvania State University, University Park, PA 16802}
\altaffiltext{28}{Institute for Gravitation and the Cosmos, The Pennsylvania State University, University Park, PA 16802}
\altaffiltext{29}{Hubble Fellow}
\altaffiltext{30}{CEA, Centre de Saclay, IRFU, 91191 Gif-sur-Yvette, France}
\altaffiltext{31}{Departamento de F\'isica e Matem\'atica, PURO/Universidade Federal Fluminense, Rua Recife s/n, Jardim Bela Vista, Rio das Ostras, RJ 28890-000, Brasil}


\date{\today}

\begin{abstract}
The Sloan Digital Sky Survey (SDSS) surveyed 14,555 square degrees, and 
delivered over a trillion pixels of imaging data. 
We present a study of galaxy clustering using 900,000 luminous galaxies with photometric redshifts, spanning 
between $z=0.45$ and $z=0.65$, constructed from the SDSS using methods described in Ross et al. (2011). 
This data-set spans 11,000 square degrees and probes a volume of  $3
h^{-3} \rm{Gpc}^3$, making it the largest volume ever used for galaxy clustering
measurements.
We describe in detail the construction of the survey window function and various systematics affecting our measurement. 
With such a large volume, high precision cosmological constraints can be obtained 
given a careful control and understanding of the observational systematics. 
We present 
a novel treatment of the observational systematics and its applications to the clustering signals from the data set. 
In this paper, we measure the angular clustering using an optimal
quadratic estimator at 4 redshift slices with an accuracy of $\sim
15\%$ with bin size of $\delta_l = 10$ on scales of the Baryon
Acoustic Oscillations (BAO) (at $\ell
\sim 40-400$ ). We also apply
corrections to the power-spectra due to systematics, and derive cosmological constraints using the full-shape of the power-spectra.
For a flat $\Lambda$CDM model, when combined with Cosmic Microwave
Background Wilkinson Microwave
Anisotropy Probe 7 (WMAP7) and  $H_0$ constraints from using 600 Cepheids observed 
by Wide Feild Camera 3 (WFC3)  (HST) , we
find $\Omega_\Lambda = 0.73 \pm 0.019$ and $H_0$ to be $70.5\pm1.6$
$ \rm{s}^{-1} \rm {Mpc} ^{-1}\rm {km}$.
For an open $\Lambda$CDM model, when combined with WMAP7 + HST, we find $\Omega_K =  0.0035 \pm 0.0054$, improved over WMAP7+HST alone by $40\%$.
For a wCDM model, when combined with WMAP7+HST+SN, we find $w = -1.071
\pm  0.078$, and $H_0$ to be $71.3 \pm 1.7$ $\rm{s}^{-1} \rm{Mpc}^{-1}
\rm{km}$, which is
competitive with the latest large scale structure constraints from
large spectroscopic surveys such as SDSS Data Release 7 (DR7) (Reid et al. 2010, Percival et al. 2010, Montesano et al. 2011) and WiggleZ (Blake et al. 2011). 
We also find that systematic-corrected power-spectra gives consistent
constraints on cosmological models when compared with pre-systematic
correction power-spectra in the angular scales of interest. 
The SDSS-III Data Release 8 (SDSS-III DR8)  Angular Clustering Data allows a wide range of investigations into the cosmological model, cosmic expansion (via BAO), Gaussianity of initial conditions and neutrino masses.
Here, we refer to our companion papers (Seo et al. 2011, de Putter et al. 2011) for further investigations using the clustering data.
Our calculation of survey selection function, systematics maps,
likelihood function for COSMOMC package will be released at
http://portal.nersc.gov/project/boss/galaxy/photoz/.
\end{abstract}



\section{Introduction}

The distribution of light in the Universe has long been used as a probe into the 
structure of the Universe. Einstein wrote of the distribution of stars as possibly 
being uniform on average over large enough distances in 1917 when he discussed 
the structure of the Universe. Hubble tested the uniformity of distribution of faint nebulae in 1926.
As the structure of the Universe unfolds, distribution of light from objects such as galaxies has remained 
a powerful cosmological probe 
\citep{peebles73,groth73,wang99,hu99,eisenstein99}.

Smoothed over large scales, we expect galaxy density to have a simple relationship
to the underlying matter density;
this implies that the clustering of galaxies at large scales is directly related
to the clustering of the underlying matter and is thus 
a sensitive
probe of both the initial conditions of the Universe and its subsequent 
evolution. 
It is therefore not surprising that a large fraction of the 
effort in observational cosmology had been devoted to measuring the spatial 
distribution of galaxies, as in the CfA Redshift Survey \citep{huchra83},The APM Galaxy Survey \citep[APM,][]{maddox90}, 
The DEEP survey\footnote{http://deep.berkeley.edu/} \citep[DEEP,][]{koo98}, 
 VIMOS-VLT Deep Survey \citep[VVDS,][]{lefevre05},
Two-Degree Field  
Galaxy Redshift Survey \citep[2dFGRS,][]{cole05},The Two Micron All sky Survey \citep[2MASS,][]{skrutskie06}, 
COSMOS \citep{scoville07}, Canada-France-Hawaii Telescope Legacy Survey\footnote{http://www.cfht.hawaii.edu/Science/} \citep[CFHTLS,][]{ilbert06}, Galaxy And Mass Assembly survey \citep[GAMA,][]{driver09}, The WiggleZ Survey \citep{blake10}, 
the Sloan Digital Sky Survey \citep[SDSS,][]{york00}.
By 2008, the SDSS\footnote{www.sdss.org} has probed $\sim1.5$ $\rm{Gpc}^3$ with galaxies, while the current SDSS-III\citep{eisenstein11} will finish surveying $\sim15$ $\rm{Gpc}^3$ in 2014. The planned Large Synoptic Survey Telescope (LSST)\footnote{http://www.lsst.org} will observe $\sim 1000$ $\rm{Gpc}^3$ of the Universe.

Hidden in the ever-increasing volume of surveyed Universe, is the wealth of cosmological information that had not been fully exploited.
In particular, the large scale clustering of any mass tracer, usually characterized
by its power-spectrum, in the Universe contains three features that are of significant interest to contemporary cosmologists.
The first distinguishing feature is oscillations in the power-spectrum caused by acoustic waves in the baryon-photon plasma before hydrogen recombination at $z\sim1000$, called Baryon Acoustic Oscillations (hereafter BAO) \citep{peebles70,sunyaev70,bond84,holtzman89,hu96,eisenstein98}.
The BAO technique has emerged as the new precision cosmology probe, especially in discerning the properties of this unknown dark component of the Universe "Dark Energy".
The BAO  was 
first observed in early 2005 both in the SDSS  
Luminous Red Galaxy sample \citep{eisenstein05},
the 2dFGRS data \citep{cole05} and in 2006 by using photometric Luminous Galaxies (LRGs) in $3500$ $\rm{deg}^2$ of SDSS \citep{padmanabhan07}.
However, neither of the signals were strong enough to place strong cosmological constraints via BAO. 
Second, the largest scales of the power-spectrum can be used to
constrain the primordial potential of the Universe, thus
testing inflation.
In particular, \cite{dalal08} has pointed out the relationship between
non-gaussianity of the potential in the early Universe (due to various possible inflationary scenarios) and the large scale power of mass tracer in the Universe.
Finally, at $k \sim 0.01 h \rm{Mpc}^{-1}$, the power spectrum turns over from a $k^{1}$ slope
(for a scale invariant spectrum of initial fluctuations), to a $k^{-3}$ spectrum, 
caused by modes that entered the horizon 
during radiation-dominated era and were therefore suppressed. The precise
position of this turnover is thus determined by the size of the horizon 
at matter-radiation equality. It corresponds to a 
physical scale determined by the total matter $(\Omega_{M}h^{2})$ densities and 
radiation densities $(\Omega_{\gamma}h^{2})$. 
In particular, with a large survey such as SDSS, various groups had used the large scale power-spectrum to put stringent constraints
on cosmological parameters, most notably \cite{zehavi02,tegmark04,eisenstein05,padmanabhan07,percival10,reid10}.

The SDSS has now surveyed $14,555$ $\rm{deg}^2$, and with appropriate photometric selection, we can construct a large 
uniform sample of the photometric luminous red galaxies \citep{ross11}, and their photometric redshifts that can be easily
calibrated using the acquired spectroscopic redshifts of a uniform sub-sample ($\sim 10\%$) of the photometric
galaxies. 
This approach allows the possibility of using both
standard rulers (from the turn over scale of power spectrum, and also the baryon acoustic oscillations) to acquire 
cosmological constraints. 

We make use of this opportunity to derive one of the most accurate
measurements of the galaxy angular power-spectra achieved to date.
We start with the five band imaging of the SDSS-III DR8; \cite{aihara11,eisenstein11}), and photometrically select a 
sample of luminous red galaxies, following the CMASS galaxy selection detailed in \cite{white11}; 
the details of the construction of the sample and the redshift distribution
is described in \cite{ross11}. We then measure the angular clustering power spectra
as a function of redshift with an optimal quadratic estimator, which is proved to provide
the best statistical error-bar when the field is Gaussian. The galaxy density field is not Gaussian on small scales, due to 
non-linear evolution; however, at relatively large scales, which are the scales we are concerned here, the field is close to Gaussian.
We will discuss this issue in detail in the paper. With such a large volume of data, we realize that the effects of large 
scale systematics are not negligible. To gauge and correct the effects of large scale systematics, we develop a novel method in 
correcting the large scale systematics given that we know the list of possible systematics.
We construct the maps of various systematics, and calculate their cross-correlation with the galaxy density, the systematic auto-correlations and cross-correlations.
We can then correct for these systematics applying this new method. 

The paper is organized as follows : Sec.~\ref{sec:sample} describes the construction
of the sample; Sec.~\ref{sec:angular} then presents the theory and measurement of the angular
power spectra; Sec.~\ref{sec:sys} discusses the various potential
systematics and the novel method applied
in correcting for the observational systematics. 
Sec.~\ref{sec:cosmo_method} describes the validation of the
cosmological parameter fitting method, and Sec.~\ref{sec:results}
summarizes the cosmological
constraints themselves.
We conclude in Sec.~\ref{sec:discuss}. 

Wherever not explicitly mentioned, we assume
best fit WMAP7 \citep{wmap7} parameters (from model lcdm+sz+lens as
specified on LAMBDA website \footnote{http://lambda.gsfc.nasa.gov/product/map/dr4/parameters.cfm}), which are: 
a flat $\Lambda$CDM cosmology with $\Omega_{M}=0.266$, $\Omega_{b}=0.0449$, $h=0.71$, 
 and $\sigma_{8}=0.801$. 

\section{The Data }
\label{sec:sample}

\subsection{Observations}

The Sloan Digital Sky Survey (SDSS; \cite{york00,eisenstein11}) mapped over a quarter of the sky using the dedicated 
Sloan Foundation 2.5 m telescope located at Apache Point Observatory in New Mexico \citep{gunn06}. A drift-scanning
mosaic CCD camera \citep{gunn98,gunn06} 
imaged the sky in five photometric band-passes \citep{fukugita96,smith02} to a limiting magnitude of r$\simeq 22.5$. The imaging data were processed through a series of pipelines that perform astrometric calibration \citep{pier03}, photometric reduction \citep{lupton00} and 
photometric calibration \citep{padmanabhan08}. In particular, Baryon
Oscillations Spectroscopic Survey (hereafter BOSS) which is a part of
SDSS III \citep{eisenstein11,aihara11}, has completed an additional
3000 square degrees of imaging and is now obtaining spectra 
of a selected subset of 1.5 million galaxies. 
The targets are assigned to spectroscopic plates (tiles) 
using an adaptive tiling algorithm based on \cite{blanton03}, and observed with a pair of fiber-fed spectrographs. 

The availability of large uniform photometric data-set prompted the start of this project, thus a series of papers, starting with the generation of the photometric redshifts catalog by \cite{ross11}, which uses 
112,778 of BOSS spectra as a training sample 
for photometric catalog. The photometric redshift catalog contains over 1.6 million objects, and 900,000 of these
objects lie within our imaging mask and the selected redshift range ($0.45$ < \rm{z} < $0.65$). 
The redshift range is selected so that it is nearly completely independent from DR7 analysis of 
LRG clustering using spectroscopy which stops at $z\sim 0.4$ \citep{reid10, percival10}; this allows
for possibility of trivial combination of likelihoods. 
These galaxies are among the most luminous galaxies in the universe and trace a large cosmological volume while having high enough number 
density to ensure shot-noise is not a dominant contributor to the clustering variance. 
The majority of the galaxies have spectral
energy distributions ($\sim 85\%$ , see \cite{masters11} and private
communication with the BOSS galaxy-evolution group) that are
distinctive of old stellar populations.

\subsection{Defining Luminous Red Galaxies}
\label{sec:lrgdef}

We make use of the CMASS sample from BOSS, which is defined in \cite{white11} and \cite{ross11}; and we write down the criteria 
here again for convenience:
\begin{eqnarray}
&& 17.5  < i_{cmod} < 19.9  \nonumber \\
&& r_{mod} -i_{mod} < 2  \nonumber \\
&& d_{\perp}  >  0.55 \nonumber \\
&& i_{fiber2} <  21.7 \nonumber \\
&& i_{cmod}  < 19.86 + 1.6 \times (d_{\perp} - 0.8) \nonumber \\
&& c_{\parallel} > 1.6 
\end{eqnarray}
where 
\begin{eqnarray}
&& d_{\perp} \equiv (r-i) - (g-r)/8 \approx r-i  \nonumber \\
&& c_{\parallel} \equiv 0.7*(g-r)+1.2*(r-i-0.18)  
\label{eq:perpdef}
\end{eqnarray}

The magnitudes denoted by ''cmod'' are ''cmodel magnitudes'' (see
\cite{white11} for more discussions), and the colors are defined with
model magnitudes, except for $i_{\rm fiber2}$, 
which is the magnitude in the $2^{''}$ spectroscopic fiber  \citep{stoughton02,abazajian04}. 
Note that we applied $i_{fiber2} <  21.7$, although the current BOSS target selection has moved the limit 
from $21.7$ to $21.5$. All magnitudes are extinction corrected using maps of \cite{schlegel98}.

In addition to constructing galaxy density maps, we created several additional maps that we use to reject regions 
heavily affected by sample systematics such as poor sky or
stellar density, and to make sure our final power-spectra are free of systematics.  
These include (i) a map of the full width at half-maximum (FWHM) of the point-spread
function (PSF) in $r$ band; (ii) a map of stellar density ($18.0<r<18.5$ stars); (iii) a map of sky brightness in $i$-band
in
nanomaggies\footnote{http://data.sdss3.org/datamodel/glossary.html\#nanomaggies}/$\rm{arcsec}^2$;
(iv) 3 map of the color offsets in $u-g$, $g-r$ and $r-i$ from
\cite{schlafly10}; (v) a map of Galactic extinction 
simply rescaled from the extinction maps from \cite{schlegel98}.

\subsection{Angular and Redshift Distributions}

To interpret the clustering of any sample, one must characterize the expected distribution
of the sample as if it is completely random. 
This involves understanding both the angular and radial selection function in addition to the expected
galaxy density, which is characterized by its mean density.

\begin{figure}
\begin{center}
\leavevmode
\includegraphics[width=3.0in]{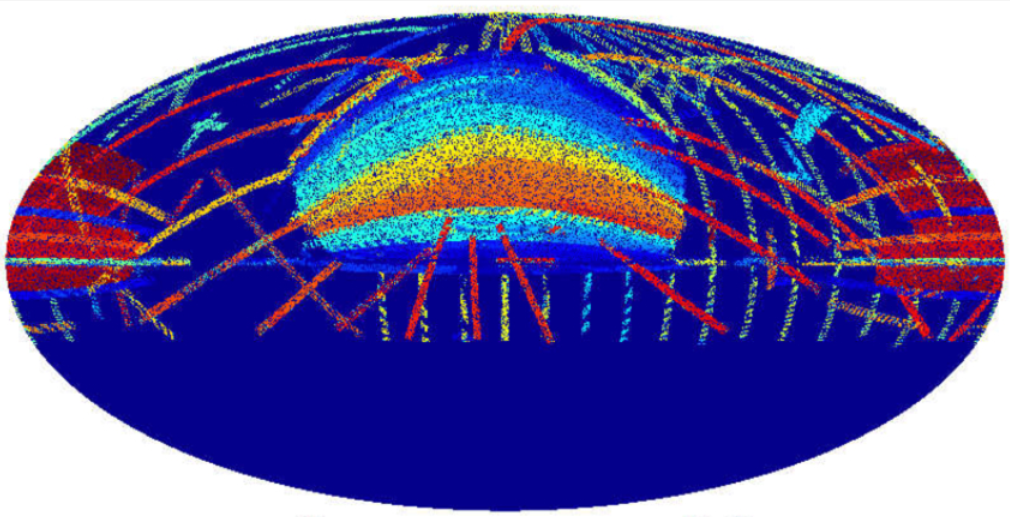}
\end{center}
\caption{The full imaging angular mask in equatorial coordinate system
  after generating a unique set of all polygons that contains primary
  fields with good observing conditions. The colors represents the
  Modified Julian Date of observation of each field.} 
\label{fig:fullmask}
\end{figure}

\begin{figure}
\begin{center}
\leavevmode
\includegraphics[width=3.0in]{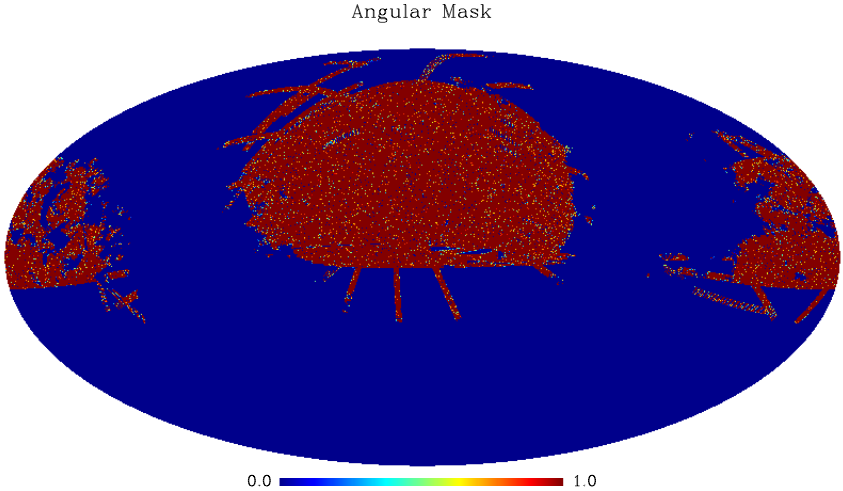}
\end{center}
\caption{The preliminary imaging mask after applying primary selection
  cuts such as cuts on seeing and the 
bright star mask on the full imaging angular mask.}
\label{fig:lrgmask}
\end{figure}

To characterize the angular window function, we generate the complete angular mask of the survey following the procedures described below.
The observed sky is defined as a union of all fields. Determining the window function requires identifying the fields that cover
each position on the sky and deciding which of those fields should be considered primary at that position. 
There is a  unique set of disjoint polygons on the sky defined by all the field boundaries, which are calculated using MANGLE package \footnote{http://spae.mit.edu/\~molly/mangle}\citep{hamilton93,hamilton04,swanson08} 
 and each field can be divided into multiple polygons.
We now must decide which fields are primary for each polygon in the
sky;  the process is described in \cite{aihara11} in detail. 
Once we determine which fields are primary for all the polygon in the sky, we make a cut on the field observing conditions (SCORE $>=$ 0.6; for more details on 
SCORE, see \cite{aihara11} or the SDSS-III webpage \footnote{sdss3.org}).
We now have an unified MANGLE polygon file that includes all the fields 
that are imaged in the entire SDSS footprint, with the correctly assigned primary fields with good observing conditions.
We call this
 the "full imaging mask", as plotted in Figure~\ref{fig:fullmask}. The color in Figure~\ref{fig:fullmask} represents the date at which the imaging 
data was taken. The striped  pattern perpendicular to the scanning
direction is easily visible, and we can also see that the north and south Galactic caps are observed 
at significantly different epoch of the survey. This provides hint as to what potential observational systematic effects would look like.
To create a more restrictive mask which is catered towards photometric red galaxies, we proceed to exclude regions where
$E(B-V)>0.08$  \citep{scranton02,ross06,padmanabhan07,ho08}, which
is almost identital to $A_r > 0.2$,  when seeing in the $i$-band exceeds $2.0''$ in FWHM, and masking regions around stars in the Tycho astrometric
catalog \citep{hog00}.
The final angular selection function covers a solid angle of $\sim
11,000$ square degrees, and 
is shown in Fig.~\ref{fig:lrgmask}.

Applying the selection criteria in Sec~\ref{sec:lrgdef} to the 14,555 $\rm{deg}^2$ of
photometric SDSS imaging considered in this paper yields a catalog
of approximately 1,500,000 galaxies. 
Applying the angular selection function as shown in Fig.~\ref{fig:lrgmask} to the Ross
et al. (2011) photometric redshift catalog yields a sample of 872,921 objects, 96\% of which are believed to be galaxies (3\% are stars, and 1\% are quasars, according 
to statistics gathered in the spectroscopic sub-sample; \cite{ross11}). For every object, photometric redshifts
and probabilities of being a galaxies were determined using the ANNz Neural Network \citep{collister04, firth03}. 
The calibration and accuracy of these data are discussed in detail in \cite{ross11}. In the range considered
in this paper,the redshifts have calibrated errors $\sim 0.04$ at $z\sim 0.45$ and $\sim 0.06$ at $z\sim 0.65$.
We pixelize these galaxies as a weighted (with the probabilities of being a galaxy) number
overdensity, $\delta_g=\delta n/\bar n$, onto a HEALPix pixelization \citep{gorski99}
of the sphere, with 12,582,912 pixels over the whole sphere (HEALPix
resolution 10, nside=1024), each pixel covers a solid angle of 11.8 $\rm{arcmin}^2$.
These pixelized maps are used directly to compute the angular power-spectra using optimal quadratic estimator.
The optimal quadratic estimator does not down-sample input pixelized maps, rather, it computes the covariance matrix
directly from these pixelized maps, and this will be discussed further in \ref{sec:est_theory}.

\begin{figure}
\begin{center}
\includegraphics[width=2.2in]{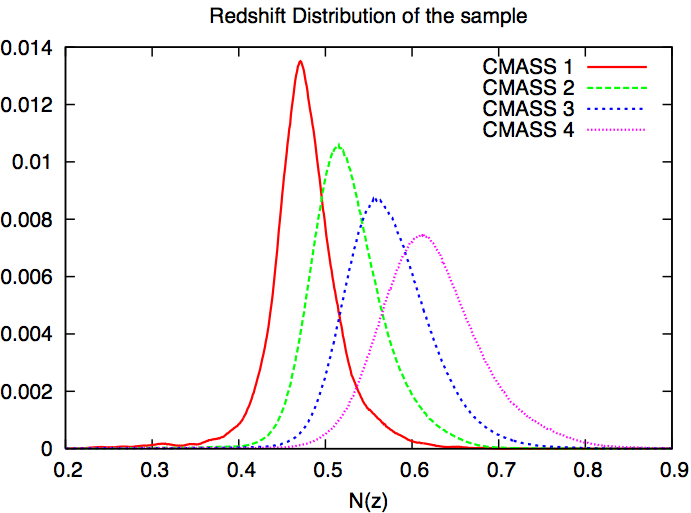}
\end{center}
\caption{The redshift distribution of the photometric CMASS sample
  when we match the objects with an unbiased sub sample from SDSS-III BOSS.}
\label{fig:dndz}
\end{figure}

\begin{figure}
\begin{center}
\includegraphics[width=3.0in]{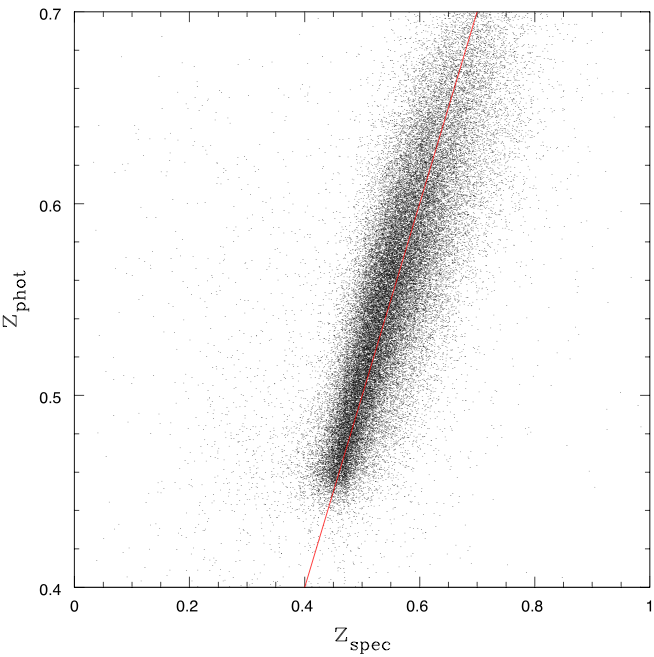}
\end{center}
\caption{The photometric vs spectroscopic redshift distribution of
  112,778 of SDSS-III BOSS CMASS galaxies.}
\label{fig:zslicecmass}
\end{figure}

The sample is divided
into 6 photometric redshift slices of thickness $\Delta z = 0.05$ starting at $z=0.4$ for CMASS sample (CMASS 0 through
CMASS 5, see Table~\ref{tab:lrgdata} for details), and the underlying redshift distributions for each slice 
are calculated using BOSS spectroscopic redshift of the same sample.
The redshift distribution of the sample is plotted in Fig.~\ref{fig:dndz}.
We can see that
although the majority of the objects in one photometric-redshift bin is
in their corresponding true redshift bin, a significant fraction of
them fall into neighboring bins. 
The comparisons of these photometric redshifts 
to the spectroscopic redshifts (obtained via SDSS III spectra) are plotted
in Fig.~\ref{fig:zslicecmass}, while properties of the different slices are summarized in 
Table~\ref{tab:lrgdata}. We see that the numbers of galaxies in both the first and the last bins are significantly smaller
than the others, therefore, we decide to drop these 2 bins from our analysis.
This decision is also facilitated by the fact that we wish to have a nearly independent sample from the 
\cite{reid10, percival10} LRG clustering analysis, thus allowing for simple combination of their likelihoods in the cosmological parameter analysis.

\begin{table}
\begin{tabular}{cccc}
\hline
Label & $z_{mid}$ & $N_{gal}$ & $b_{g}$ \\
\hline
\input{lrgdata.tbl}

\hline
\end{tabular}
\caption{\label{tab:lrgdata} Descriptions of the 6 $\Delta z=0.05$ redshift slices;
$z_{mid}$ is the midpoint of the redshift interval. Bias parameters are deduced from marginalizing over all the other cosmological parameters (and a free shot noise term)
from combining WMAP 7 + HST + DR8 angular power-spectra likelihood using only $ 30< \ell <150$ multipoles. The first and last bins are dropped
from here on due to the small number of galaxies in those bins.}
\end{table}

\subsection{Sample Systematics}

There are a number of potential systematic effects in photometric samples that
contaminate clustering: stellar contamination and obsuration,
seeing variations, sky brightness variations, extinction
and color offsets (such as those described in \cite{schlafly10}). 
\cite{ross11} had extensive discussions on these potential systematics; 
we will concentrate on the particular effects from various systematics on the angular power-spectra in the range of scales that affects our science analysis. 

The above cuts remove only parts of the sky that are significantly affected by extinction and seeing variations.
With such a large sky coverage, an accurate determination of the angular power-spectra of the 
the large scale tracer is only possible through an thorough understanding of the systematics.
However, if we only retain parts of the sky that have the minimum
systematics effects; we must remove most of 
the coverage, as we have demonstrated in \cite{ross11}.
Therefore, we developed a novel way of dealing with residual sample systematics which we will discuss in Sec.~\ref{sec:theory_sys} in detail.

\section{The Angular Power Spectrum}
\label{sec:angular}

As was noted in the introduction, the angular power-spectrum contains information of both the 
growth and the expansion of the Universe through two standard rulers of the Universe: 
the baryon acoustic oscillations and 
the matter-radiation equality turn-over scale; thus the shape of the power-spectrum. 
In this section, we will describe both the theory and the computation of angular power spectrum. 

\subsection{From galaxy distributions to angular power-spectrum}
\label{sec:theory1}

The intrinsic angular galaxy fluctuations are given by:
\begin{equation}
g(\thetaB)= \int dz \,b(z) N(z) \delta(\chi(z)\thetaB,z),
\label{dg}
\end{equation}
where $b(z)$ is an assumed scale-independent bias factor relating the galaxy overdensity to
the mass overdensity, i.e. $\delta_g =b\,\delta$,
$N(z)$ is the normalized selection function, and $\chi(z)$ is the comoving distance to redshift $z$.
We focus on the auto power-spectrum of the galaxies:
\begin{equation}
\label{cc1}
C^{gg}(\ell)=\frac{2}{\pi} \int k^2 dk P(k) [g]_\ell (k) [g]_\ell(k)
\end{equation}
where $P(k) = P(k,z=0)$ is the matter power spectrum today as a function of the wave number $k$,
and the function $[g]_\ell$ is
\begin{equation}
\label{cc2}
\left[g\right]_\ell(k)=\int dz \, b_i(z) N(z) D(z)j_\ell(k\chi(z))
\end{equation}

The Limber approximation, which is quite accurate when $\ell$ is not
too small ($\ell \gsim 10$),
can be obtained from Eq.~(\ref{cc1}) by setting $P(k) = P(k=(\ell+1/2)/\chi(z))$ and using the
asymptotic formula that $(2/\pi)\int k^2 dk j_\ell (k\chi) j_\ell (k\chi') = (1/\chi^2) \delta (\chi-\chi')$ (when $\ell \gg 1$). We find that the
substitution
$k=(\ell+1/2)/\chi(z)$ is a better approximation to the exact expressions than $k=\ell/\chi(z)$.
Note that $j_{l}(x)$ is the $l^{\rm th}$ order spherical Bessel function.  
On large scales where the mass fluctuation $\delta \ll 1$, the
perturbations grow according to linear theory
$\delta(k, z)$ = $\delta(k,0) D(z)/D(0)$.

For auto-correlation, applying Limber approximation will change Eq.~\ref{cc1} to the following: 
\begin{equation} 
C^{gg}_\ell=\int dz \frac{1}{\chi^2(z)} b^2(z) N^2(z) P(k,z)
\label{cc2}
\end{equation}

For cross-correlation between two different large scale structure samples (be it different selection functions, redshift
distributions, different biases), we can write the cross-correlation as follows:
\begin{equation} 
C^{gg'}_\ell= \int dz \frac{1}{\chi^2(z)} b(z) b'(z) N (z) N^{'}(z)  P(k,z)
\end{equation}
where $g'$ can have different biases, redshift dependence etc. 

We have not yet distinguished between the galaxy and the matter angular power-spectrum yet. 
Throughout this paper, we simply assume
\begin{equation}
\label{eq:non-linear_eq}
C_{g}(\ell) = b_{g}^{2} C_(\ell) + N_{\rm{shot}} + a  \,\,,
\end{equation}
where $C_{g}(\ell)$ and $C_(\ell)$ are the galaxy and matter angular
power spectra; 
$b_{g}$ is the linear galaxy bias, $N_{\rm{shot}}$ is a constant shot
noise term which is estimated by the optimal quadratic estimator and $a$ is a constant term that is fitted as a freely floating parameter.
This is a good approximation on large scales,
but breaks down on smaller
scales; we defer a discussion of its regime of validity, as well as the nonlinear
evolution of the power spectrum to a later section of this paper \ref{sec:non_linear}.

Throughout the paper, we adopt this linear redshift independent (within our redshift slice) bias model with a constant shot noise term. The bias and the shot noise term of galaxy 
sample for the various 
redshift slices are fit as extra parameters in Cosmological Monte
Carlo (COSMOMC; \cite{cosmomc}) chains to ensure we do not bias our cosmological models via fixing any particular pre-computed bias.

\subsection{Redshift-Space Distortions}
\label{sec:theory2}

The position of observed galaxies can be inferred from their redshift, and hence the peculiar 
velocity along the line-of-sight can in principle affect our angular power-spectrum.
So far we have neglected the effect of the peculiar velocity, i.e., the redshift-space distortion 
(RSD) effect on the angular power spectrum. 
In the 3-dimensional redshift-space power spectrum measured with spectroscopic surveys, 
the modeling of RSD is still challenging due to the fact that the mapping process 
from real to redshift-space is nonlinear in terms of peculiar velocity. 
For recent efforts, see for example \cite{Scoccimarro:2004tg,Taruya:2010mx,Reid:2011ar,Seljak:2011tx}.
It is comparatively easy to model the RSD effect on the angular power spectrum, 
because the RSD information along the line of sight is projected out in the angular clustering. 
\cite{padmanabhan07} formulated the RSD for the angular power spectrum
at the linear level, 
and showed that the linear RSD effect can be seen only at large scales ($\ell<20$). 
However we could imagine that, if we select thin redshift slices, the nonlinear RSD 
effect may not be projected out and becomes non-negligible at small scales. 
\cite{Saito:2012xx} shows that such nonlinearities becomes important
only in the case when
$\sigma_{z}<0.01$ at $\ell>500$ but this is not the case here. 

We here include the linear RSD effect following \cite{padmanabhan07}. 
To be complete, let us review some of the important details from \cite{padmanabhan07}. 
\be
1+g(\thetaB) = \int \, d\chi\,N(s)\,[1+\delta(\chi\thetaB,\chi)]\,,
\label{eq:deltag_red}
\ee
where we have now written the normalized selection function as a function of redshift-space 
distance, $s = \chi + {\bf v}\cdot\thetaB$ with the peculiar velocity component, ${\bf v}$. 
Assuming the peculiar velocities are small compared with the thickness of the redshift slice, 
we Taylor expand the selection function to linear order, 
\be
  N(s) \approx  N(\chi) + \frac{dN}{d\chi}({\mathbf v}\cdot\thetaB)\,.
\ee
Substituting this expression into Eq.~\ref{eq:deltag_red}, 
we express separately the 2D galaxy density field in two terms, 
$g = g^{0} + g^{r}$, where $g^{0}$ is the term discussed in the previous section, 
while $g^{r}$ is the linear RSD correction. 
With the help of linear continuity equation, we have the Legendre coefficient as 
\be
\delta_{g}^{r}(\ell) = i^{\ell}\,\int \frac{d^{3}k}{(2\pi)^{3}}\,W_{\ell}^{r}(k)\,. 
\ee
The component is given by 
\be
W_{\ell}^{r}(k) = \frac{\beta}{k}\int\,d\chi\frac{dN}{d\chi}\,j'_{\ell}(k\chi)\, , 
\ee
where $\beta$ is the growth parameter defined by $\beta \equiv d\ln D/d\ln a /b_{g}$, and 
$j'_{\ell}$ is the derivative of the spherical Bessel function with respect to its argument. 
We can then apply the fact that $C_{\ell} \equiv \langle g_{\ell} g_{\ell}^{*} \rangle$, and
calculate the redshift space distorted angular power-spectra.

\subsection{Non-linearities}
\label{sec:non_linear}

Non-linearities in the power-spectrum are caused by the non-linear evolution of components of the Universe, especially 
the late time evolution of matter and baryons. To capture the full extent of the non-linearities, with a lack of 
full-fledged non-linear evolution theory, one will need to 
simulate the evolution of most if not all of the components of the Universe. Extensive research and discussion have been 
carried out in multiple fronts \citep{sanchez08,sanchez09}, whether it is by perturbation theory \citep{carlson09}, dark matter simulations
\citep{hamaus10, heitmann09}, or fitting functions suggested by dark matter simulations \citep{smith03}. 
Historically, there are a few ways to deal with non-linearities in utilizing power-spectrum to constrain cosmology,
such as comparing the non-linear power-spectrum to the linear power-spectrum (usually for specific cosmological model),
and keeping only scales that are believed to be linear \citep{tegmark04, padmanabhan07}; or utilizing the halo occupation model to 
convert a galaxy power-spectrum into a halo power-spectrum, which can be easily compared to halo power-spectra from 
dark matter simulations\citep{reid10}; or using a variety of fitting functions developed \cite{carlson09} to fit its 
observed galaxy power-spectra \citep{blake10}. 
Our project both benefits and suffers from the fact that it is a photometric survey. On one hand, 
its BAO signal is smeared as we don't have accurate redshifts; on the other hand, the integration along lines of 
sight ameliorates the non-linearities that would have been considerably stronger. 
Therefore, traditionally, angular power-spectra analysis usually only applies a simple cut on the angular scale that 
roughly corresponds to $k= 0.1k \rm{Mpc}^{-1}$ \citep{padmanabhan07}. In this paper, we take a small step forward in terms of 
non-linearity treatment of the overall shape of angular power-spectrum, and also adopt a similar treatment as in \cite{eisenstein07,blake10} for the non-linear treatment on the BAO scales.

\subsubsection{Non-linear effects on the overall shape of the power-spectrum}

There is an extensive literature discussing how one can model the linearities of 3D power-spectrum over a large range of 
scales \citep{sanchez08,carlson09, hamaus10}. This paper does not intend to address the issue 
of fully modeling the non-linearities in 3D power-spectrum; we do, however, take a simple model that happens 
to perform quite satisfactorily for the 2D angular power-spectrum.
We adopt the simple linear redshift-independent biasing model (with shot noise subtracted for every single angular power-spectra). 
Therefore, in addition to the cosmological parameters that are of
interest for each model, we include three extra parameters for each 
redshift slice ($b$, $N_{\rm{shot}}$ and $a$) as shown in \ref{eq:non-linear_eq}.

We test the sufficiency of this model in multiple ways. 
We test this model by fitting only $2$ < $\ell$ < $150$ and $2$ < $\ell$ < $200$ using simulated CMASS mocks (as is discussed in Sec.\ref{sec:simulations}).
We compute optimally quadratic estimated power-spectra of simulated data (a total of 160 realizations 
from 20 independent simulation boxes, 8 lines of sight each), and then we compute 8 averaged (over 20 independent simulations) power-spectra,
and combined it with a pseudo-WMAP7 likelihood (which has the covariances of WMAP7 likelihood, but with cosmological
parameters 
centered on the input parameters of the CMASS mocks. 
We find that when using above mentioned model for the averaged power-spectra, in combination with pseudo-WMAP7, 
we recover {\it all} input cosmological parameters of the CMASS mocks
for all 8 averaged power-spectra to within $1.5 \sigma$. We conclude that a spread over $1.5 \sigma$ is reasonable. 
The bias parameters recovered are also similar to the input bias of
the CMASS mocks as described in \cite{white11}. We therefore conclude that this model is accurate in recovering cosmological parameters when used in the range of angular scales as specified above.

\begin{figure}
\begin{center}
\leavevmode
\includegraphics[width=3.0in]{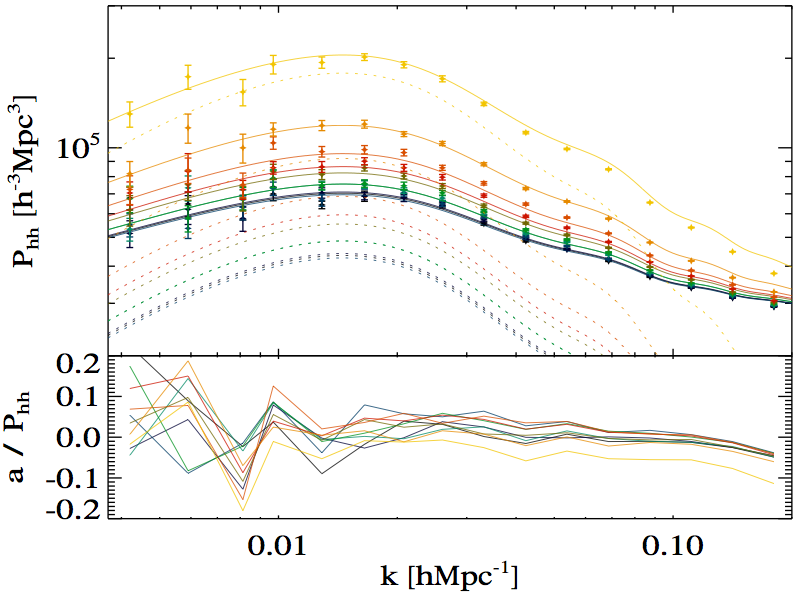}
\end{center}
\caption{ To justify our choice of scale for fitting our cosmological parameters, and the model we adopted, we
show how well simple model such as $b^2 P_{\rm halofit}(k) +1/n + a $ can fit fairly well up to $k=0.2 h/Mpc$. Top panel shows
the non-linear power-spectrum of halos in a cosmological simulations (dots with error bars), the 
model being considered here $b^2 P_{\rm halofit}(k) +1/n $ (solid line) and the dashed line shows what happens if we only use $b^2(k) P_{\rm NL}(k)$ instead. Bottom
panel shows $a/P_{\rm hh}(k)$ where
 $a= P_{\rm halofit}(k) +1/n$, we can see that the ratio is fairly consistent with 0.0 for a large range of mass, and it starts to
deviate from
0.0 starting at $k=0.1 $Mpc/h. Therefore, we find it prudent to
include an extra parameter a in our formalism, as we do include some
modes at k larger than $k=0.1$ Mpc/h. 
The different color lines correspond to different halo mass ranges. The largest halos are those with highest bias, which 
also gives the largest deviations from the model.}
\label{fig:Pk}
\end{figure}

We further tests this model by comparing this model with \cite{hamaus10}, we found that our simple method fits the non-linear
power-spectrum derived from cosmological simulations quite well even up to $k=0.2 h/Mpc$. In Figure~\ref{fig:Pk}, we plot the non-linear power-spectrum from numerical simulations of halos (points with errorbars), while the solid lines are power-spectrum of various halo mass bins calculated using 
our simple model $b^2 P_{\mathrm non-lin}(k) + \frac{1}{\bar{n}}$ ,
the model fits the non-linear power-spectrum quite well over a significant range
in k even when we have not yet added the additional constant term a. The dashed lines show the results without the shot noise term for
various halo mass bins. 
Our model of non-linear power-spectrum is based on HALOFIT \citep{smith03}, so in order to not confuse the reader, we will 
call $P_{\mathrm non-lin}(k)$ by $P_{\rm halofit}(k)$.
The lower panel shows the ratios between $a \equiv P_{\rm hh}(k) - (b^2 P_{\rm halofit}(k) +1/n )$ and $P_{hh}(k)$ are plotted as lines. 
The non-linear bias is fairly well fit by our simple
model even if we do not include the extra constant bias term. We
decided to include the extra constant term a to help remove the residual difference 
between $P_{\rm hh}(k)$ and $b^2 P_{\rm halofit}(k) +1/n$. 

\subsubsection{Non-linear effects on the BAO}

We test the effect on our results of non-linear evolution on the
smearing of the BAO feature by assuming that the non-linear matter power spectrum follows the expression in \cite{eisenstein07}:
\begin{eqnarray}
P(k)&=&\exp\left(-k^2\Sigma_{\mathrm nl}^2/2\right)P_{\mathrm wiggle}(k)+ \nonumber \\
 &&\left(1-\exp\left(-k^2\Sigma_{\mathrm nl}^2/2\right)\right)P_{\mathrm no-wiggle}(k)
\end{eqnarray}
where $\Sigma_{\mathrm nl}=7.527h^{-1}$Mpc, $P_{\mathrm wiggle}(k)$ is
the linear theory power-spectrum (which includes the BAO) and 
$P_{\rm no-wiggle}(k)$ is a smooth power spectrum,
with the same shape as $P_{\mathrm wiggle}(k)$ but without any baryonic oscillations;
which is computed using the approximation
described in \cite{eisenstein98}. 
Both the wiggle and the no-wiggle part have been computed in linear
theory; we then added to both of them the corresponding nonlinear
ratios as a function of the scale. This approach significantly enhances the
power in small scales. We find that the results are not very sensitive
to the exact value of $\Sigma_{\mathrm nl}$ provided that it is in the
 range of 5.527 to 9.527 $h^{-1}$Mpc \citep{eisenstein07}.
In principle, $\Sigma_{nl}$ is cosmology dependent, and thus can
change our cosmological constraint if it is kept as a free parameter. 
We have therefore examined our constraints on cosmological parameters 
using different
$\Sigma_{nl}$. We test this issue by fitting the full set of cosmological parameters
using MCMC fitting method with COSMOMC with $\Sigma_{nl}$ set to $2 h \rm{Mpc}^{-1}$ higher 
and lower than its currently chosen value (7.527), and find that when
we fit for a $\Lambda$CDM model in combination with WMAP7, there is less than $5\%$ change for 
any of the parameters.

\begin{figure}
\begin{center}
\leavevmode
\includegraphics[width=3.0in]{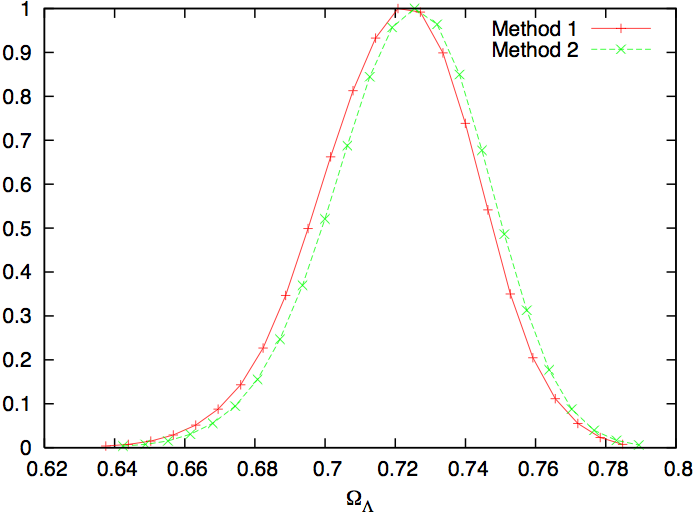}
\end{center}
\caption{We applied two different methods in calculating the power-spectrum (including the BAO) with non-linear effects taken into account, and find that it makes very little difference in the cosmological parameter constraints.}
\label{fig:method12}
\end{figure}

However, the addition of the nonlinear ratios is quite important, not only because the power in small scales in the angular power spectrum at high multipoles is not expected to be accounted for the shot noise due to finite number of galaxies, but also because the small shift in the BAO wiggles can slightly modify the best-fit shape of the power spectrum and hence return a different value of $\Gamma \equiv \Omega_m h$. We applied two different methods in calculating the 
power-spectrum (including the BAO) with non-linear effects taken into
account, and find that it makes essentially no difference (see Figure~\ref{fig:method12}).

\subsection{Optimal Estimation of Angular Power Spectrum}
\label{sec:est_theory}

The theory behind optimal power spectrum estimation is now well established,
so we limit ourselves to details specific to this discussion, and refer
the reader to the numerous references on the subject
\citep[][and references therein]{hamilton97,seljak98}.
We also refer the reader to
the Appendix \ref{sec:quad_imple} for more specific details that relate 
to our paper directly.

We start by parametrizing the power spectrum with twenty step functions in
$l$, $\tilde{C}^{i}_{l}$,
\be
C_{\ell} = \sum_{i} p_{i} \tilde{C}^{i}_{\ell} \,\,,
\ee
where the $p_{i}$ are the parameters that determine the power spectrum. We
form quadratic combinations of the data,
\be
q_{i} = \frac{1}{2} {\mathbf x}^{T} {\mathbf C}_{i} {\mathbf C}^{-1}
 {\mathbf C}_{i} {\mathbf x}\,\,,
\ee
where ${\mathbf x}$ is a vector of pixelized galaxy overdensities,
${\mathbf C}$ is the covariance matrix of the data, and ${\mathbf C}_{i}$
is the derivative of the covariance matrix with respect to $p_{i}$.
The covariance matrix requires a prior power spectrum to account for
cosmic variance; we estimate the prior by computing an estimate of the
power spectrum with a flat prior and then iterating once.
We also construct the Fisher matrix,
\be
F_{ij} = \frac{1}{2}
  {\rm tr} \left[{\mathbf C}_{i} {\mathbf C}^{-1} {\mathbf C}_{j}
{\mathbf C}^{-1}\right] \,\,.
\ee
The power spectrum can then be estimated, $\hat{\mathbf p} =
{\mathbf F}^{-1} {\mathbf q}$, with covariance matrix ${\mathbf F}^{-1}$.

We also refer the reader to Appendix \ref{sec:quad_imple} for details more specific to
our project.

\subsection{Tests with simulations}
\label{sec:simulations}

\begin{figure}
\begin{center}
\leavevmode
\includegraphics[width=3.2in]{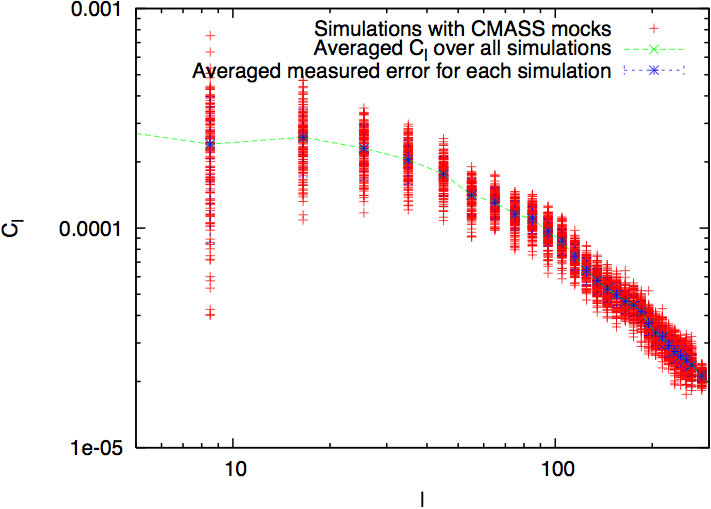}
\includegraphics[width=3.2in]{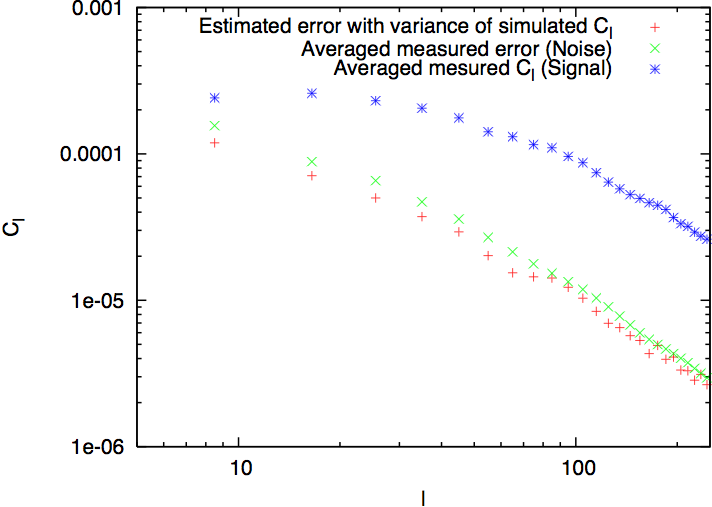}
\includegraphics[width=3.2in]{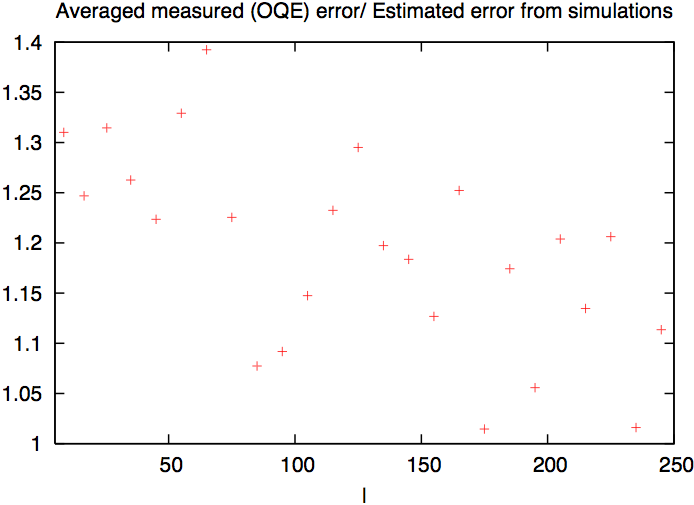}
\end{center}
\caption{(top)The estimated (from Optimal quadratic estimator) power
  spectrum from 160 simulated realizations (red crosses) out of 20
  independent dark matter only simulations \citep{white11}. 
The green points show the averaged recovered power-spectrum from each of the
simulations, while the blue points show the averaged measured 
error as estimated from optimal quadratic estimator. 
We conclude that the averaged measured error from the optimal
quadratic estimator is a good measure of the expected error.
(middle) The estimated error (red crosses) is based upon the variance of the estimated power-spectrum from each simulation at each $ell$-mode. We have also 
plotted the averaged measured error (green points), and it is a bit
higher than the estimated error from the variance of the simulations
(bottom panel). Nevertheless, we show that the
quadratic estimator code can estimate the errors of the power-spectrum
in the scale of interest here with reasonably high accuracy.}
\label{fig:plotsims}
\end{figure}

To test whether the errors estimated by the 
quadratic estimator employed here are accurate or not and to test the
results of our pipeline, we
must compute the errors obtained via a series of simulations. 

One way to do this is to generate Gaussian random field using the prior power spectra 
for each redshift slice to simulate 
over the entire sphere. We can Poisson distribute galaxies 
with probability $(1+\delta)/2$ over the survey region, trimmed with the angular selection
function. 
\cite{padmanabhan07} has tested this pipeline with the Gaussian random fields simulations, thus
what we need to test here is whether the errors estimated by the
quadratic estimator are appropriate, considering
that the power-spectrum measurement is only minimum variance measurement when the field is Gaussian, which is not 
the case here. Given the non-gaussianity of the field, we need to determine how close we are to minimum variance measurement.

As we would like to simulate our galaxy sample as closely as possible, we employed CMASS mock catalogs from \cite{white11} to test the accuracy of the optimal quadratic estimator. 
\cite{white11} has produced a series of mock catalogs that use the
best-fit HOD models from \cite{white11}, and populate a series of
N-body simulations \citep{white11}. 
The majority of the galaxies are central galaxies living in halos of mass $10^{13}  h^{-1}M_{\sun}$.
We generate 8 lines of sight from each corner of each of the 20
independent CMASS simulations from \cite{white11}.
These mock catalogs are then processed the same manner as the real data through the quadratic estimator
code, and analyzed in the same manner as the real data set. The mock angular power-spectra are
thus optimally estimated angular power-spectra. 

We plotted the distribution of the power-spectrum from each simulation
that are estimated by the quadratic estimator code, and compare 
these results to the averaged error-bar of the simulation (see Figure~\ref{fig:plotsims}). 
When comparing the expected error to the distribution of estimated power-spectrum from each simulation and the averaged measured error from each simulation,
we conclude that the averaged measured error is a good measure of the expected error.
We have plotted the estimated error (red crosses of the middle panel
of Figure ~\ref{fig:plotsims}) by examining the variance of the estimated power-spectrum from each simulation at each $\ell$-mode. We have also
plotted the averaged measured error (green points), and it is a bit higher than the estimated error from the variance of the simulations (bottom panel). 
This is probably due to the fact
that since there are only 20 simulation boxes, with 8 lines-of-sights overlapping slightly in within each box. Therefore, the variance of simulated $C_l$ is probably
slightly smaller than it should be at all scales, due to the correlations between lines of sight. Regardless, we show that the quadratic estimator code can estimate the errors of the power-spectrum in the scale of interest here
with reasonably high accuracy.
It is important to note that at all scales of interest (to the current paper), the estimated error from the quadratic estimator code is not under-estimated.

\subsection{The Optimally Estimated Angular Power Spectrum}
\label{sec:angpower}

\begin{figure}
\begin{center}
\includegraphics[width=3.0in]{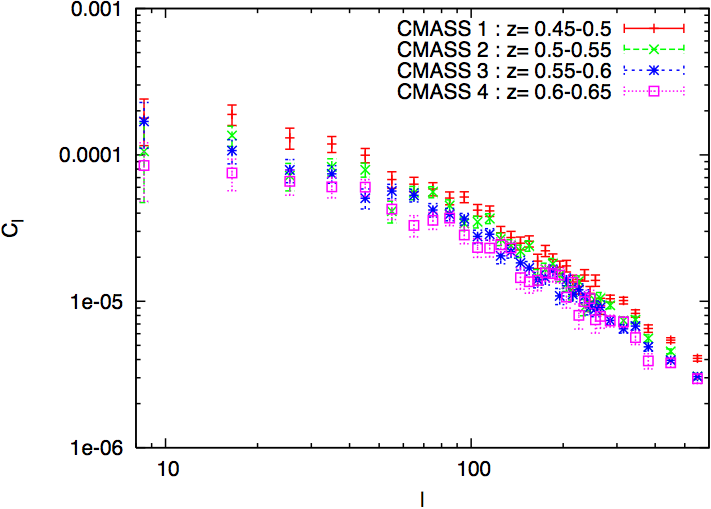}
\includegraphics[width=3.0in]{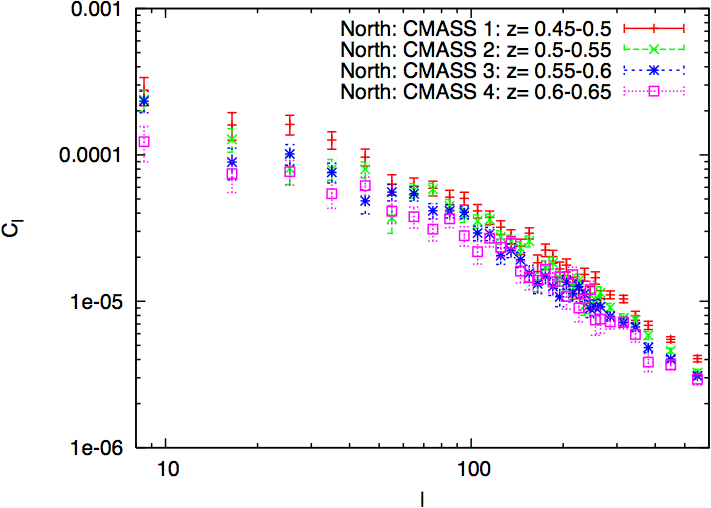}
\includegraphics[width=3.0in]{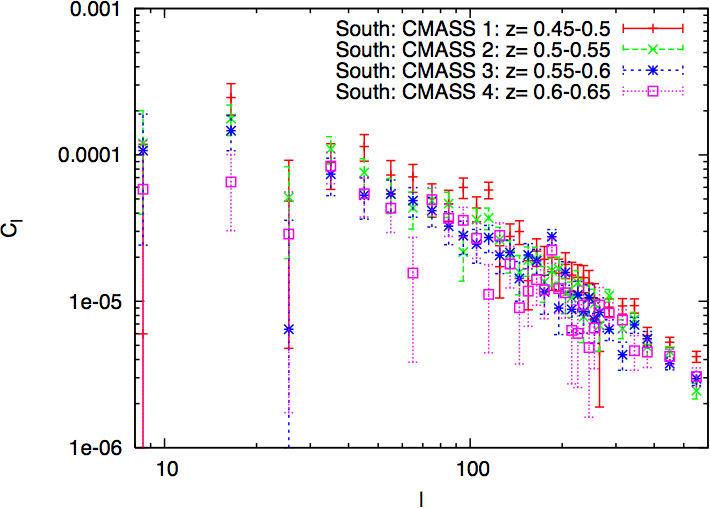}
\end{center}
\caption{The measured angular power spectrum for the 4 redshift bins using methodology described in Sec~\ref{sec:est_theory}. 
We have plotted the full angular power-spectra, which takes into the whole sky in the top panel, the north (Galactic) angular
power-spectra and the south (Galactic) angular power-spectra. Within the range of interest, the north and south angular power-spectra are consistent, suggesting that the systematics
are at a relatively low level in the scales of interest, if they affect the north and south differently.}
\label{fig:cl2dfine}
\end{figure}

The angular power-spectra estimated using the methodology described in Sec~\ref{sec:est_theory}
are displayed in Figure~\ref{fig:cl2dfine}. 
In particular, we plot separately the north (Galactic), south and full angular power-spectra of these 4 redshift bins (CMASS 1-4, from $z=0.45-0.65$). 
We plotted the north and south separately to investigate possible systematic differences due to the long separation of 
observation time between north and south galactic caps. For the scales
of interest ($30< \ell < 150$),
the north and south are not different 
to prompt separate analyses. Nevertheless, this does not
preclude possibility of systematic differences at the largest scales
(at low multipole) in
the angular power-spectrum. This is only possible, since the estimated
power in each $\ell$-bin is not correlated, therefore a systematic
difference in one $\ell$-bin does not affect another.

To test the similarity of north and south region on scales of interest
( $30< \ell < 150$), we find all best fit cosmology parameters (with
combined with WMAP7, via MCMC chain using COSMOMC) found
by north and south alone respectively are consistent with each other. 
It is interesting to note that the south has smaller area than the north, and thus there are less 
information per l-bin, thus the error-bars in the south is significantly larger than the north. It will also be discussed later in Sec~\ref{sec:sys} as the systematic treatment presented in this paper
will in principle correct systematic variations even when the full
survey is analyzed in one piece. 
We can also see the evolution of the angular power-spectra over different redshift slices, as it is expected.

\begin{figure}
\begin{center}
\includegraphics[width=3.0in]{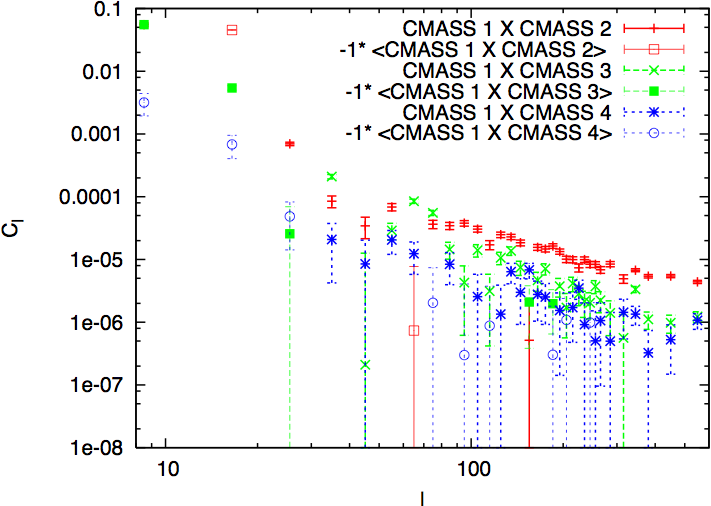}
\includegraphics[width=3.0in]{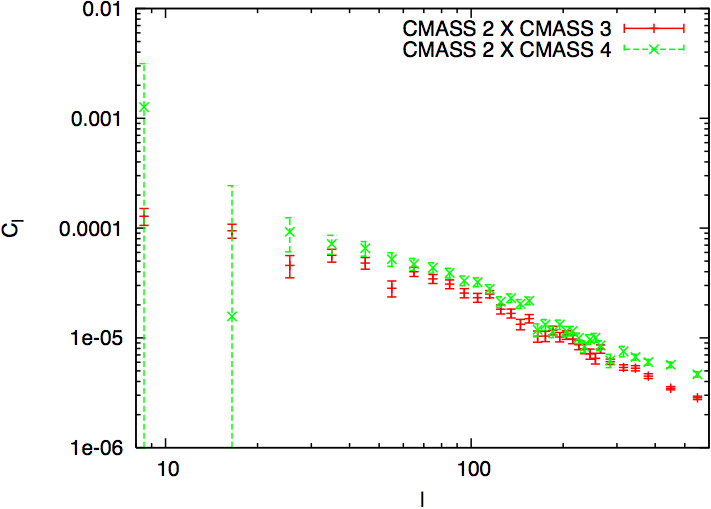}
\includegraphics[width=3.0in]{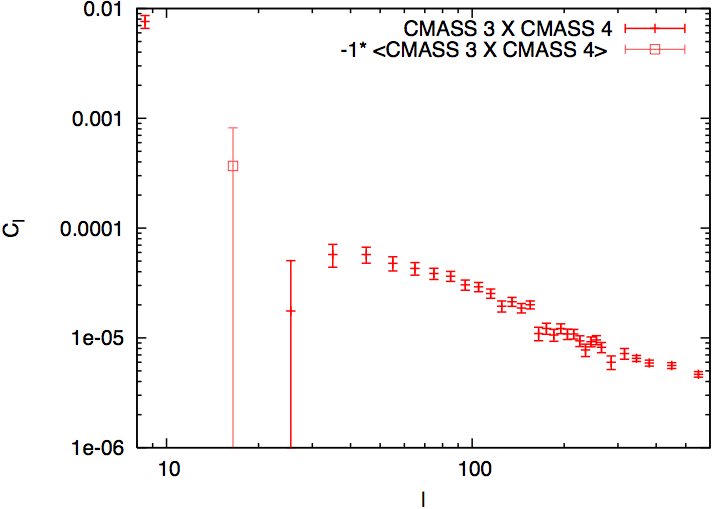}
\end{center}
\caption{The measured angular cross-correlations for the first 3 redshift bins with the other slices. 
We do not show repeats of the cross-correlated pairs. When we examine cross-power across various redshift bins, any 
difference between the measured power and the expected power (from galaxy cross-correlations) can also be used as a measure of 
the effects of systematics. In the top panel, there is significant extra power at large scale, and also negative correlations (which cannot 
come from galaxy auto-correlations), therefore, there are significant
systematics within CMASS 1. In the bottom panel, we observe the 
high redshift slice CMASS 4 also has some substantial effects from systematics at large scales. The CMASS 2 and CMASS 3 samples 
are fairly clean from systematics in the scales of interest.
}
\label{fig:clcross}
\end{figure}

As shown in Figure~\ref{fig:dndz}, we need to investigate the potential effects of overlapping redshift distributions. 
We calculate the cross-power of various redshift combinations, and they are shown in Figure~\ref{fig:clcross}.  
Cross-power between different redshift bins not only add information in terms of cosmology, but also from the perspective of systematics.

When we examine cross-power across various redshift bins, any
difference between the measured power and the expected power (from galaxy auto-correlations in the same 
redshift range) can be used as a measure of 
the effects of systematics. In the top panel, there is significant extra power at large scale, and also negative correlations (which cannot
come from galaxy auto-correlations), therefore, we know that there is
significant systematics within CMASS 1. The bottom panel shows that the
high redshift slice CMASS 4 also has substantial effects from systematics.

Finally, to estimate whether it is worth including the cross-power of
various redshift slices into the cosmological analysis, we performed a simple Fisher analysis.
We calculated Fisher matrices using angular spectra from the four redshift
bins (CMASS 1-4), with the redshift distributions given in Figure~\ref{fig:dndz}.
A standard $\Lambda$CDM cosmology is employed to calculate the fiducial
spectra. We used the Limber approximation (where the input power spectrum
was given by CAMB \footnote{http://camb.info/} linear power-spectrum and HALOFIT) and ignored redshift space
distortions.
We employ the standard Gaussian expression for the covariance matrix of the spectra.
The shot noise term was calculated assuming $N_l$ = $1/\bar{n}$ (with
$\bar{n}$ being the number of galaxies per steradian of the individual bin).
Finally, to construct the Fisher matrix, we used the range $l=30-300$. The
parameter space is given by: $\Omega_b$,$\Omega_c$, $\Omega_\nu$, $\Omega_\Lambda$, $\sigma_8$, $n_s$, $b_1$, $b_2$, $b_3$ and $b_4$ ($b_N$ refers to biases of galaxy sample at redshift slice N).
The Fisher matrix is then added to WMAP7 Fisher matrix
and invert to find the covariance matrix for the
parameters.
We then consider two cases: (1) using only the auto-spectra as observables
and (2) using both auto- and cross-spectra as observables. The errors on
all parameters improve by less than 5\% in going from (1) to (2). We also
found that ignoring covariances between different auto-spectra 
(we do include the covariance between auto power-spectra in the analysis) makes less than 5\% difference.
This suggests that when we include these covariances in the MCMC, the
errors will not change significantly.
We therefore adopt a conservative approach where we don't include the cross-power as extra signal, but 
we include the bin-to-bin covariance that can, in principle, be double-counted due to the 
overlap of redshift slices.

\section{Potential Sample Systematics}
\label{sec:sys}

Without accurately addressing known potential systematics on the observed number density of objects in our sample,
we cannot claim to understand its 
expected angular power-spectra, nor can we extract cosmological information from it. 
The treatment of systematics is especially crucial for the overall shape of power-spectrum, since 
the shape does not deviate much from power-laws, and has no specific features such as those in BAO.
The oscillatory nature of the BAO signal helps it from being contaminated by any systematic signal 
that 
doesn't have oscillatory features. Moreover, most BAO detection methods attempt to minimize any influence
from the shape directly \citep{eisenstein07}, thus further shielding
the BAO technique from any systematic effects. 
We will propose a novel method of systematic corrections in
Sec~\ref{sec:theory_sys} which helps mitigate the effects in power-spectrum caused by any
systematics. 

\subsection{Description of Systematics}

Here we consider not only sample systematics, but in particular the systematics
that may contribute to extra (or deficit) power in the angular scale
under consideration. 

\subsubsection{Stellar contamination and obscuration} 

\begin{figure}
\begin{center}
\includegraphics[width=3.0in]{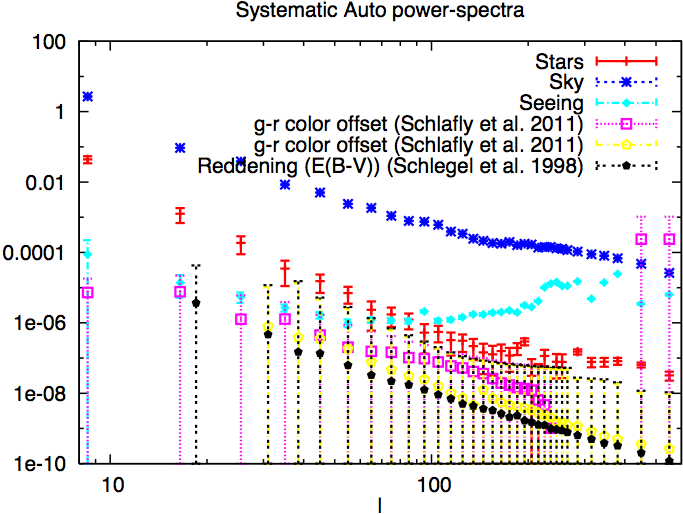}
\end{center}
\caption{The angular power-spectra of various systematics we investigated in relation to the 
possible contamination to the galaxy power-spectra} 
\label{fig:sys_power}
\end{figure}

\begin{figure}
\begin{center}
\includegraphics[width=3.0in]{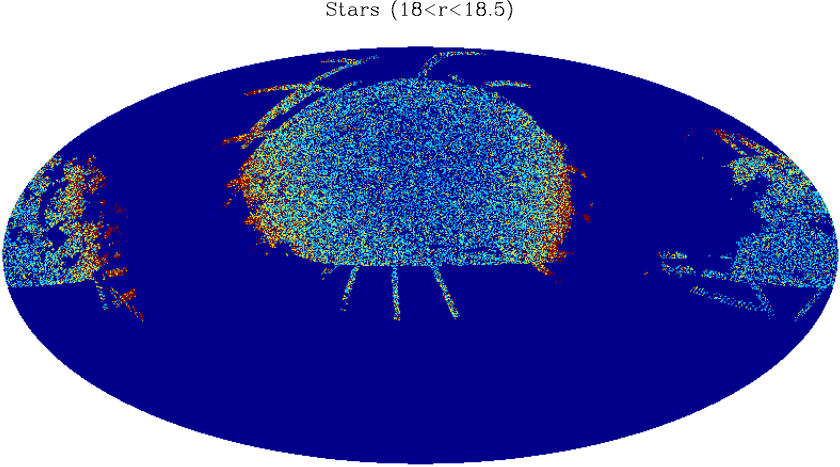}
\end{center}
\caption{ The stellar density map constructed from stars of $18<r<18.5$.}
\label{fig:star_map}
\end{figure}

\begin{figure}
\begin{center}
\includegraphics[width=3.0in]{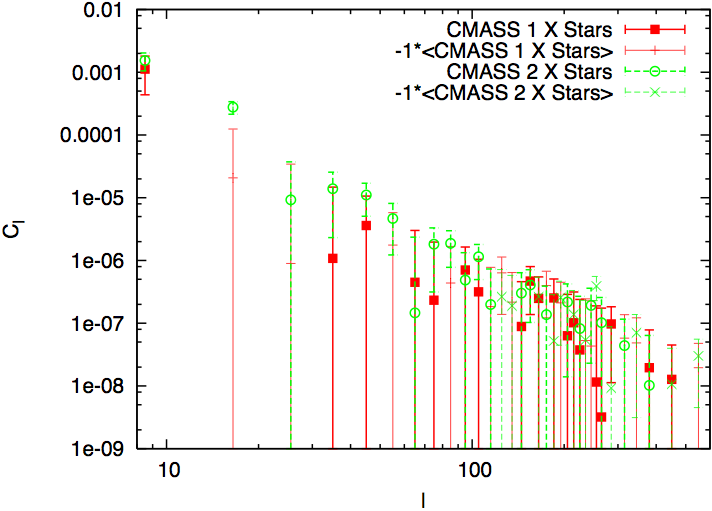}
\includegraphics[width=3.0in]{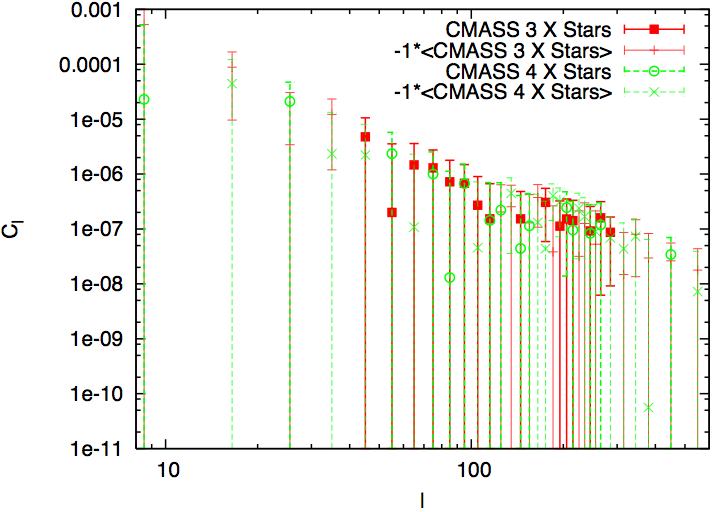}
\end{center}
\caption{The cross-correlations between the various galaxy overdensity slices and stellar overdensities of $18<r<18.5$.}
\label{fig:star_cross}
\end{figure}

Stars can in principle mimic galaxies given the right colors, or give rise to obscuration due to possible foreground subtraction issues due to presence of
a star. 
As it was pointed out in \cite{ross11} that the magnitude range of stars do not change its effect
on the galaxy number density, we pick stars of magnitude $18<r<18.5$, and investigate its influence
on the galaxy auto power-spectrum.
The stellar density map is plotted in Figure~\ref{fig:star_map}, while its auto power-spectrum
is plotted in Figure~\ref{fig:sys_power}. We use the same mask as the CMASS samples, since the stars
can only affect the galaxy power-spectrum where the two overlap. 
We calculate the cross-power-spectra between the stars and the various CMASS redshift slices, 
and find that there is a significant correlation at several angular scales (see Fig~\ref{fig:star_cross}), mostly at large scales. 
In particular, there are strong angular correlations ($\ell < 10$ for
CMASS 1, $\ell < 20$ for CMASS 2) between stars and the galaxies at
large scales, while we observe
the number of density of galaxies is lower when it is closer to a star
(as also discussed in \cite{ross11}).In the paper, we do not include
scales that are smaller than $\ell < 8$ since it is much larger than the scales we are interested in this paper. 
However, we will discuss more for the larger scales in a future publication on primordial non-gaussianities as the largest angular scales contain more information concerning primordial non-gaussianities.

There is an extensive discussion on the stellar contamination in the
CMASS catalog in \cite{ross11}.The fundamental conclusions are that there are two separate effects: 1) stars can be confused as galaxies, thereby contaminating the sample and inducing a positive correlation between the densities of stars and our sample and 2) the presence of a star artificially reduces the chances of detecting a galaxy, thereby imparting a negative correlation. In \cite{ross11}, since the band powers are highly correlated across
bins of separating distances, the two effects together impart a slightly negative correlation between the number density of stars and our sample. 
In our analysis detailed in this paper, the estimated angular power-spectra are designed to have minimal correlation across bins, therefore, 
we can see both positive and negative correlations over different scales ($\ell$-bin), as seen in Figure.~\ref{fig:star_cross}.
Given that we know stars are likely to contribute to the observed number densities,
we can take into account of the amount
of contamination by using the above discussed technique. 
Our results are consistent with
\cite{ross11} even though we do not detect smaller scale correlations between the stars 
and the galaxies, since the estimator employed in this paper produces estimates of angular band powers that are minimally correlated with other bins of band powers, while estimates are highly correlated across bins in 
the analysis of \cite{ross11}. Therefore, although the correlations
between stars and galaxies concentrated in the largest scales, they
appear in smaller angular scale such as those seen in \cite{ross11}.

\subsubsection{Sky brightness} 

\begin{figure}
\begin{center}
\includegraphics[width=3.0in]{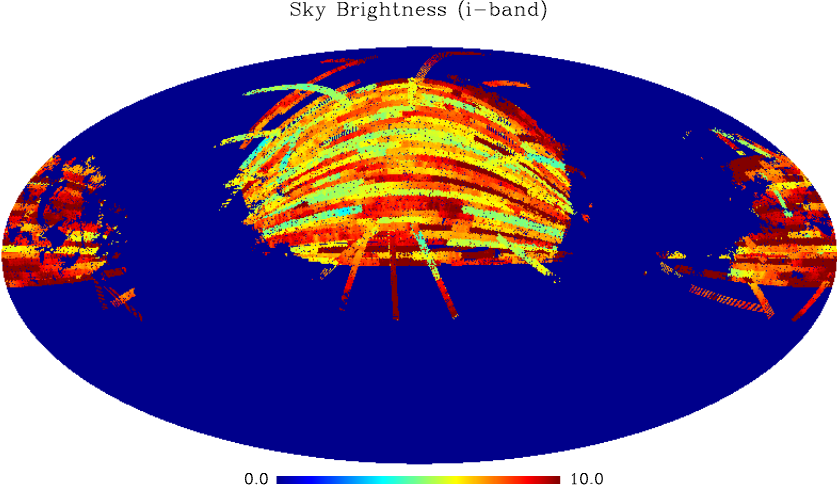}
\end{center}
\caption{ The sky brightness characterized by the $i$-band sky brightness in nano-maggies/$\rm{arcsec}^2$.}
\label{fig:sky_map}
\end{figure}

The sky brightness from SDSS is presented in
Figure~\ref{fig:sky_map}. There are at least two scales that the sky
brightness would affect the 
expected number of galaxies. 
The first is the width of the scan of the SDSS camera, 
the second comes from the fact that the southern cap has brighter sky,
since we observe the south at higher zenith angle, thus more
re-emission from the optically thin lines that were pumped by the sun originally. 

\begin{figure}
\begin{center}
\includegraphics[width=3.0in]{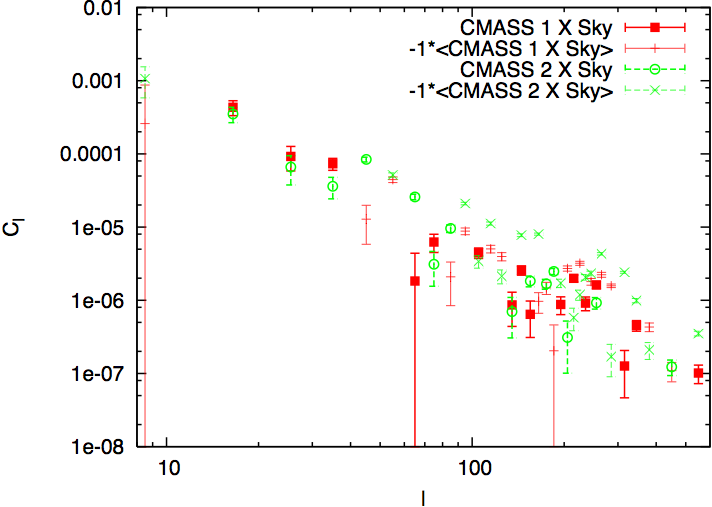}
\includegraphics[width=3.0in]{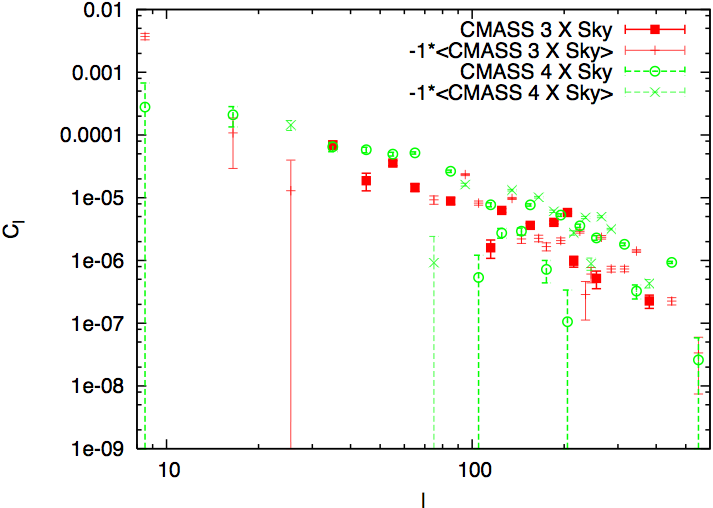}
\end{center}
\caption{ The cross correlations between sky brightness ($i$-band) and galaxy overdensities at various redshifts.}
\label{fig:sky_cross}
\end{figure}

We use the auto power-spectrum of the sky brightness (shown in Figure~\ref{fig:sys_power}), and its cross-correlations
with galaxy densities at various redshifts (shown in Figure~\ref{fig:sky_cross}), to estimate the amount of contamination that 
can come from the sky. We discuss the corrections applied arising from
sky brightness in Section~\ref{sec:theory_sys}.

\subsubsection{Seeing Variation} 

\begin{figure}
\begin{center}
\includegraphics[width=3.0in]{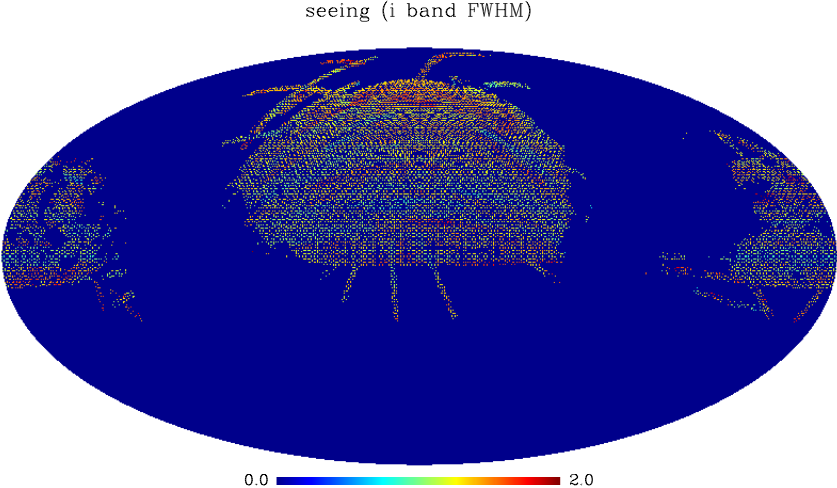}
\end{center}
\caption{Seeing (image quality) map plotted in $i$-band Full-Width Half Maximum (FWHM). The stripes are probably
caused by the changing atmosphere seeing over the observation time, as
different parts of the sky are observed on different nights.}
\label{fig:seeing_map}
\end{figure}

Since SDSS uses a ground-based telescope at Apache Point Observatory,
it is expected that the image quality, primarily atmosphere seeing
will affect the number of galaxies detected in any part of the sky. To quantify this, we plotted the seeing variations in the sky in Figure~\ref{fig:seeing_map}. 
There is a striped pattern as different parts of the sky are observed
in different nights, which have different atmosphere seeing.  
We use the auto power-spectrum of the seeing variations and its cross-correlations with the galaxy density to determine the effects of seeing on the 
galaxy overdensity clustering power. Since we can see that there are statistically significant but mild cross-correlations between the galaxies and
seeing in several of angular band power, we correct for the 
seeing variations as discussed in Section~\ref{sec:theory_sys}

\begin{figure}
\begin{center}
\includegraphics[width=3.0in]{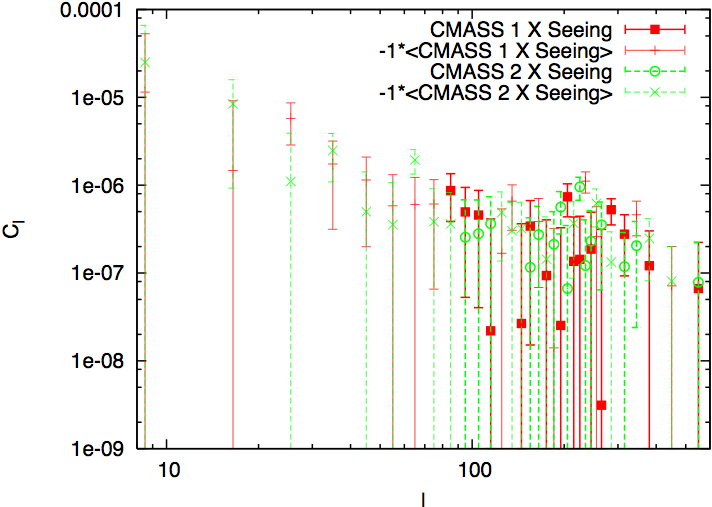}
\includegraphics[width=3.0in]{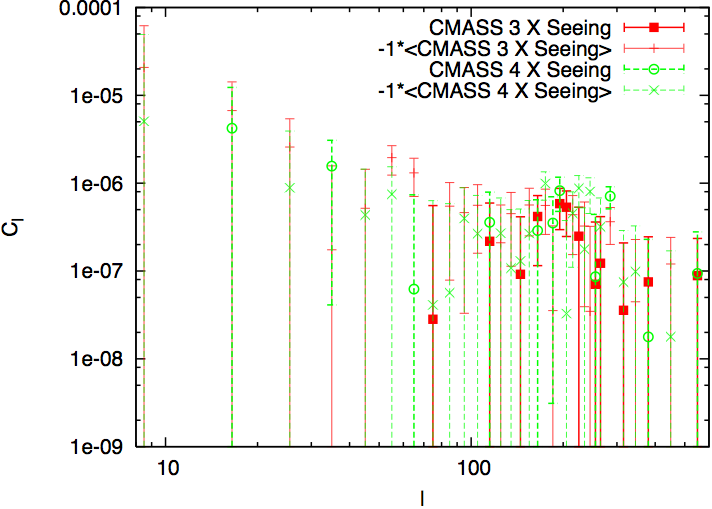}
\end{center}
\caption{ The cross correlations between image quality (seeing) and galaxy overdensities at various redshifts. } 
\label{fig:seeing_cross}
\end{figure}

\subsubsection{Extinction} 

\begin{figure}
\begin{center}
\includegraphics[width=3.0in]{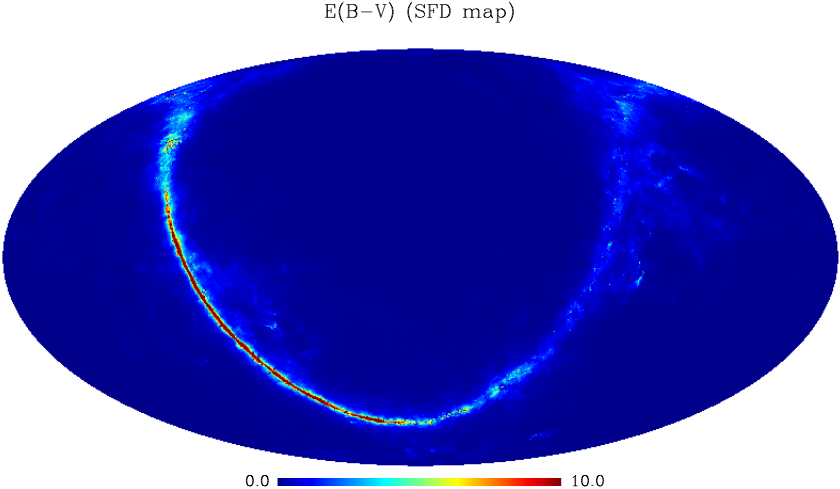}
\end{center}
\caption{The Galactic extinction map from \cite{schlegel98}. By
  comparing with the full mask of the sky, we can see that regions of
  maximum extinction, the Galactic plane is completely avoided.}
\label{fig:dust_map}
\end{figure}

We check for any residual effects on the observed galaxy
over-densities due to Galactic extinction by computing the
cross-correlations between the galaxy overdensities and the extinction
map \citep{schlegel98} (see Figure~\ref{fig:dust_map}). 
Since SDSS avoids most heavily extincted areas, we only have a small overlapping area where there is significant extinction, and galaxy data.
We do not see a statistically significant cross-correlation between the galaxy (except for scale of $\ell$ < $20$  of CMASS 1) and the extinction field, therefore, 
we conclude that we will drop the galactic extinction from the list of
possibly contributing systematic effects as long as the range of interest in
this analysis remain smaller than $\ell$ > $20$.

\begin{figure}
\begin{center}
\leavevmode
\includegraphics[width=3.0in]{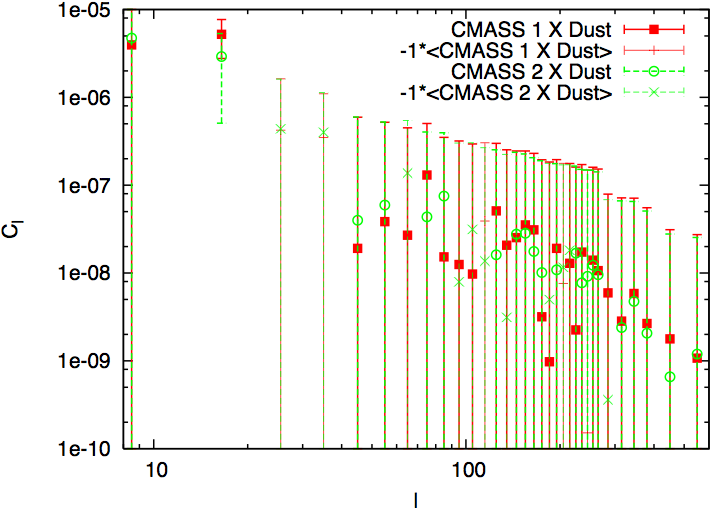}
\includegraphics[width=3.0in]{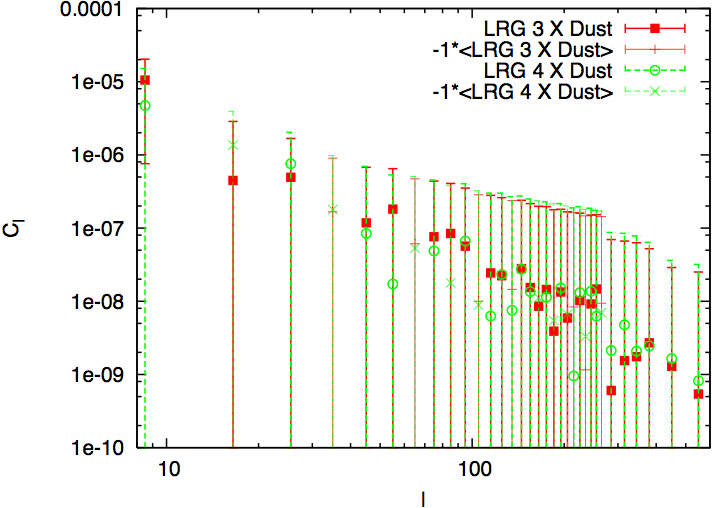}
\end{center}
\caption{The cross-correlations between galaxy of various
  redshift slices with the Galactic extinction map. Since there are
  not significant correlations between the galaxies overdensities and
  the Galactic extinction map, we can drop the extinction from the
  list of potential systematic effects for CMASS 1, CMASS 3 and CMASS
  4 . The CMASS 2 sample has a significant contribution at 1
  multipole, but for reasons that we will discuss in later sections,
  we would not include multipoles at $\ell < 30$, so this one multipole at CMASS 2 would not affect
our cosmological analysis. }
\label{fig:dust_cross}
\end{figure}

\subsubsection{Color offsets} 

\cite{schlafly10} reported various color offsets for the SDSS
footprint, in particular a north/south offset.
As discussed in \cite{schlafly10}, the photometric offsets can be
estimated via two different ways:  1) using the color of stars in the
imaging data; 2) using the stellar spectra to determines spectral
classes, and then calculate differences between the observed and
expected colors of stars. 
We adopted the later method, since it will not be sensitive to the intrinsic variations of stellar properties. 
However, this approach requires spectroscopy of stars, which is
lacking in significant parts of SDSS southern sky. We still pursues it though, and found that there are no significant detection between the galaxy density map that overlaps with the offset map (which is lacking in southern sky coverage). We therefore conclude that this is not 
an important systematics in our sample. 
However, this is only a statement that at the sky which are overlapped between the galaxy density map and the offset map, there isn't significant correlations. We will need more data in terms of the southern sky offsets before we can conclude on the effects of color offsets and galaxy densities.

\begin{figure}
\begin{center}
\leavevmode
\includegraphics[width=3.0in]{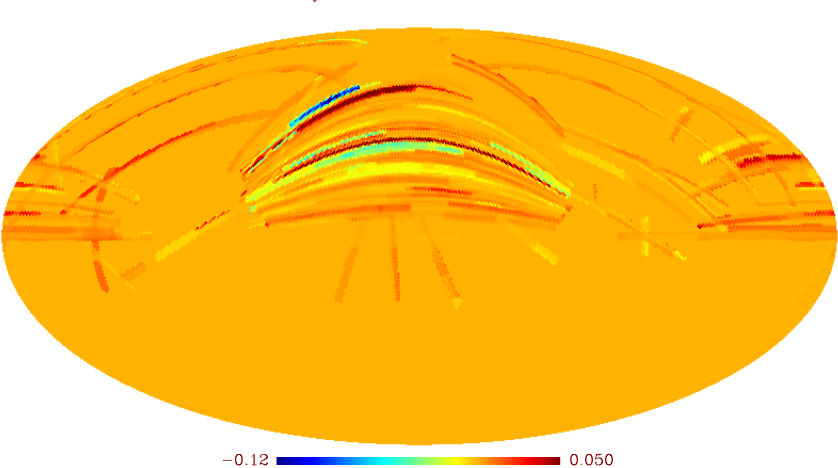}
\end{center}
\caption{The color offsets in  $r$-$i$ calculated from sources provided by Schlafly et al. (2010). These are derived from the spectra of the stars, thus does not include an intrinsic variations due 
to the metallicity gradient of stars within our galaxy.}
\label{fig:schlafly_map}
\end{figure}

\begin{figure}
\begin{center}
\leavevmode
\includegraphics[width=3.0in]{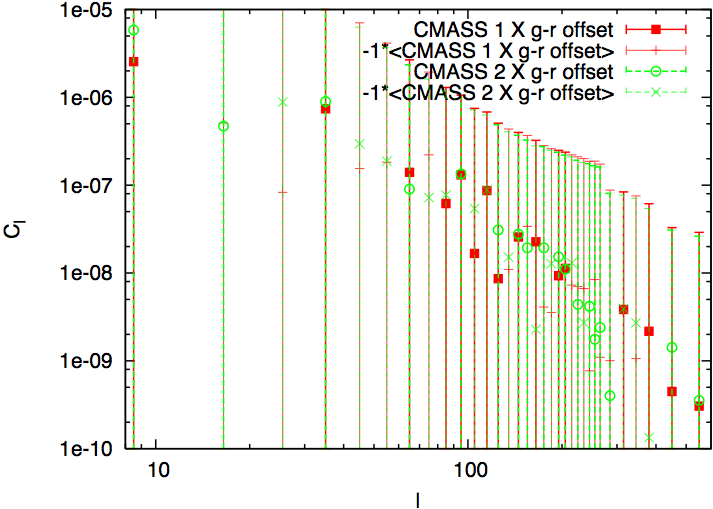}
\includegraphics[width=3.0in]{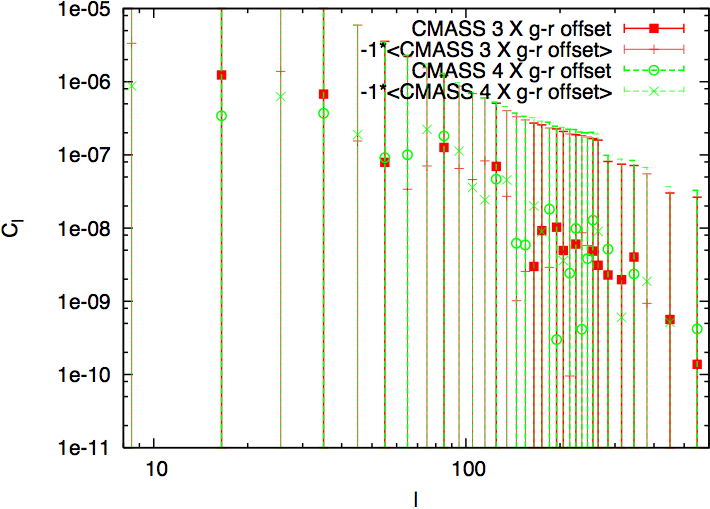}
\includegraphics[width=3.0in]{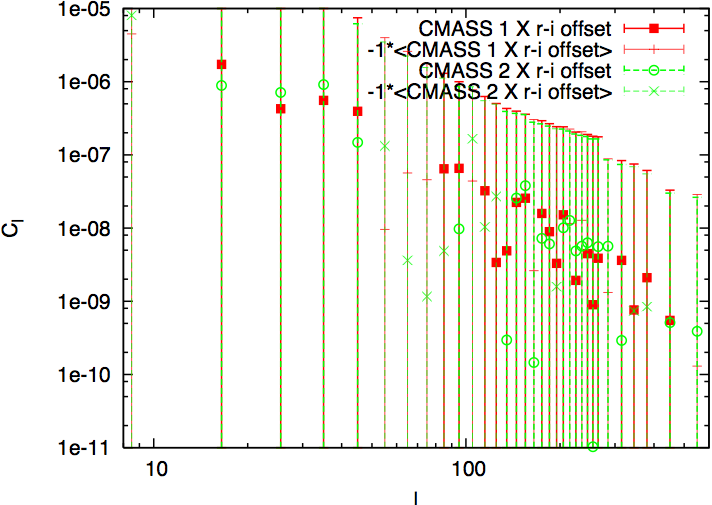}
\includegraphics[width=3.0in]{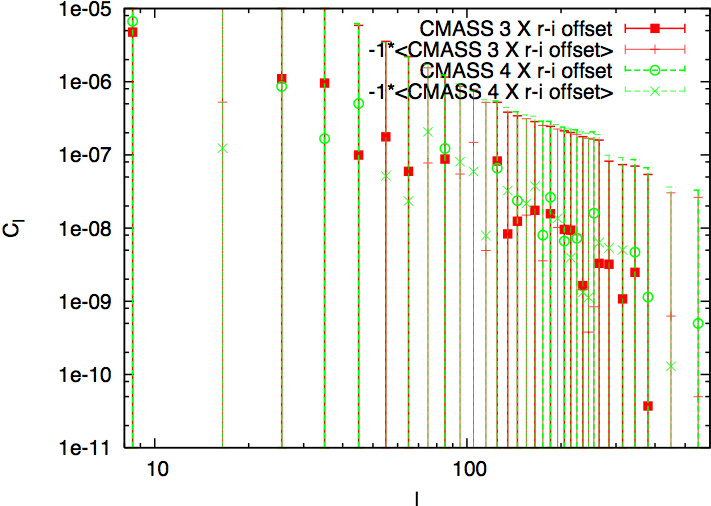}
\end{center}
\caption{ The cross-correlations between the galaxy overdensities and the color offsets in g-r (top two panels) and r-i (bottom two panels). We can see that there is no significant correlation between the galaxies and the color offsets, removing these offsets from our list of potential systematics}. 
\label{fig:schlafly_cross}
\end{figure}

\subsubsection{Magnitude Errors} 

As mentioned in \cite{ross11}, the model magnitude errors in the southern cap are larger than the northern cap by $\sim 10\%$, 
this may introduce a possible power excess (or deficit) at the lowest multipole. 
This issue however shouldn't affect any of the other multipoles which are
the focus of our paper.

\section{Novel Treatment of Systematics}
\label{sec:theory_sys}

Assuming that residual systematics will exist best catalog we can
construct without
a serious loss of sky,
how should we handle the remaining systematics?
Without any evidence of possible non-linear effects of systematics on the observed density fields, we adopt the 
simplest approach: linear relationship between the systematics and the observed galaxy density fields.

We start from the following:
Transforming fields from real space into spherical harmonic space (or l space in particular), so that $<\delta_g \delta_g> = C_l$:

\begin{equation}
\delta_g^o = \delta_g^{t} + \sum_{i=1}^{N_{sys}} \epsilon_i \delta_i
\end{equation}
\noindent
where $\delta_g^t$ is the true galaxy density (in the $\ell$-space), and each $\delta_i$ is the fluctuations of
the map of the i-th systematic, while $\epsilon_i$ characterizes how much
i-th systematic contributes.
With the lack of a better model, we assumes a linear relationship
between the systematics and the observed galaxy number overdensity, but
in principle this model can be modified to include higher order
contamination due to the systematics.

For a simple demonstration, we consider that we have only two systematics contributing to the observed
galaxy density, so that $i=2$. Assuming that the true galaxy density
is unrelated to any of our
systematics, we have the following:
\begin{equation}
\left \langle \delta_g^o \delta_g^o \right \rangle = \left \langle \delta_g^t \delta_g^t \right \rangle +
\epsilon_1^2 \left \langle \delta_1 \delta_1 \right \rangle +
\epsilon_2^2 \left \langle \delta_2 \delta_2 \right \rangle +
2 \epsilon_1 \epsilon_2 \left \langle \delta_1 \delta_2 \right \rangle
\end{equation}

Furthermore, we have all the cross-correlations between the systematic and the observed galaxy density map:
\be
\left \langle \delta_g^o \delta_1\right \rangle = \epsilon_1 \left \langle \delta_1 \delta_1 \right \rangle +
\epsilon_2 \left \langle \delta_1 \delta_2 \right \rangle
\ee
and
\be
\left \langle \delta_g^o \delta_2\right \rangle = \epsilon_2 \left \langle \delta_2 \delta_2 \right \rangle +
\epsilon_1 \left \langle \delta_1 \delta_2 \right \rangle
\ee

Since we calculate all the auto- and cross- correlations of all the systematics (on top of the cross-correlations between the systematics and the observed galaxy density), we can solve for
$\epsilon_1$ and $\epsilon_2$ (and they will be functions of $\ell$).

\begin{figure}
\begin{center}
\leavevmode
\includegraphics[width=3.0in]{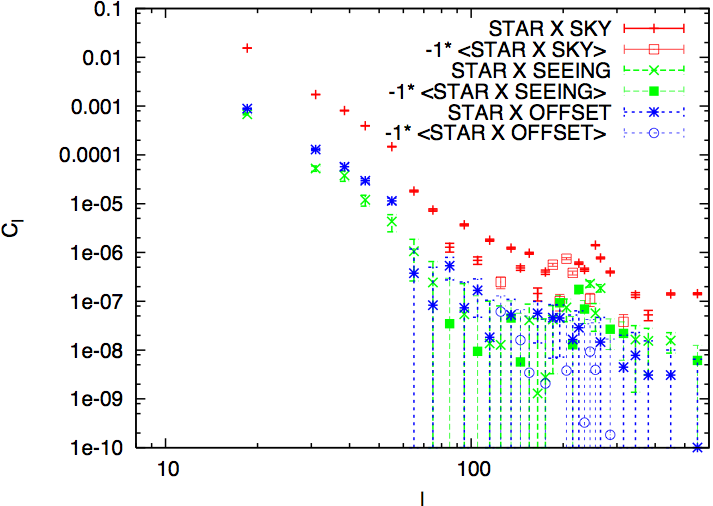}
\includegraphics[width=3.0in]{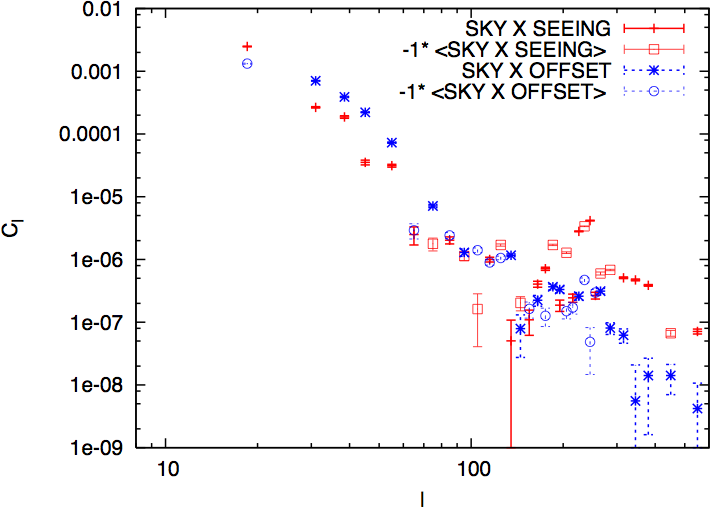}
\includegraphics[width=3.0in]{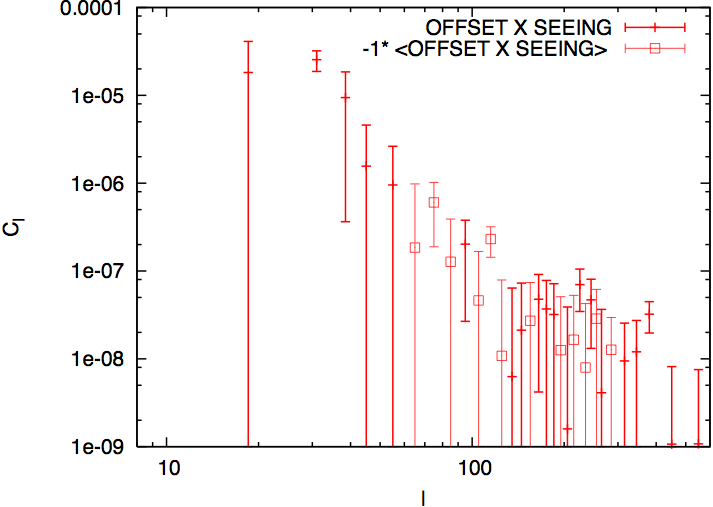}
\end{center}
\caption{ The cross-power among systematics which are found to
  contaminate observed galaxy field. In particular, we show the
  cross-correlations between stellar contaminations (STARS), sky
  brightness (SKY) and image quality (SEEING). }
\label{fig:syscross}
\end{figure}

In Figure~\ref{fig:syscross}, we show the cross-correlations of all
the systematics (which contaminates the observed galaxy field),  we find that the correlations 
across different systematics are far from zero, and we must include
the cross-correlations among systematics in our model.

\begin{figure}
\begin{center}
\leavevmode
\includegraphics[width=3.0in]{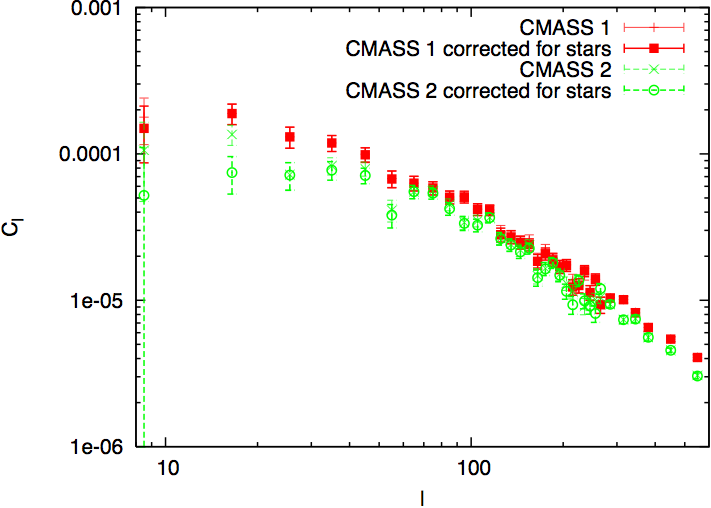}
\includegraphics[width=3.0in]{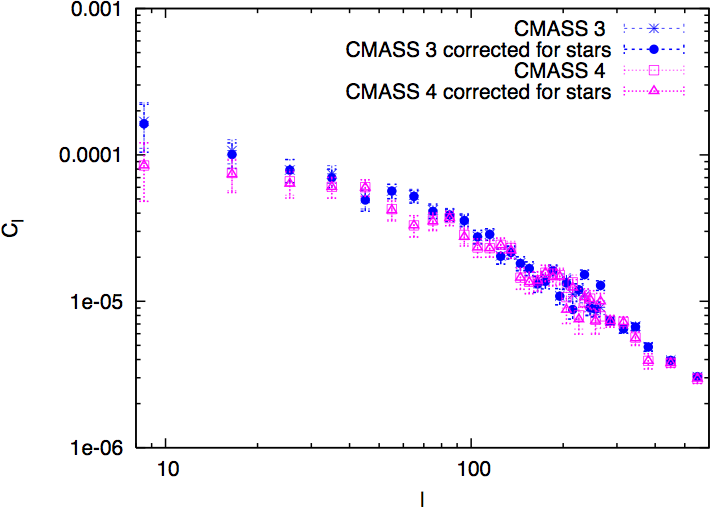}
\end{center}
\caption{We show the corrections made by inclusion of only 1
  systematics (stellar contamination). We can see that there are
  significant corrections in large scales, but the systematic does not affect the
  smaller scales. }
\label{fig:starsyscor}
\end{figure}

For simplicity of demonstration, we show the application of applying the correction from only 1 systematic (stars) in Figure~\ref{fig:starsyscor}. Since there is only inclusion of 
1 systematic, the correction depends only on the auto-power of the stars and the cross-correlation between the stars and the observed density field. 
Although star is one of our dominant systematics, its effect on
scales of interest ($\ell$ > 30) are quite minimal. This implies that
with the appropriate estimator (which do not correlate powers in
various scales), effects from systematics can be corrected relatively
easily.
This result further encourages us in terms of the cosmological
constraining power that can be harnessed from future imaging surveys
that will go deeper and wider.

\begin{figure}
\begin{center}
\leavevmode
\includegraphics[width=3.0in]{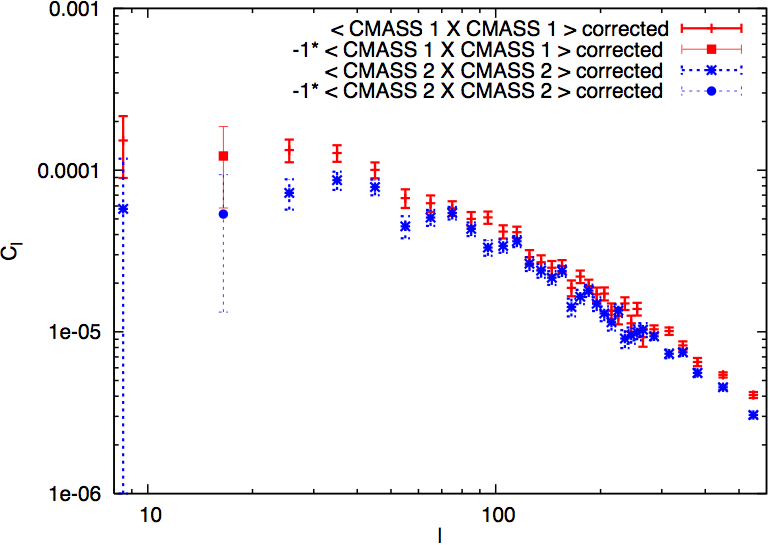}
\includegraphics[width=3.0in]{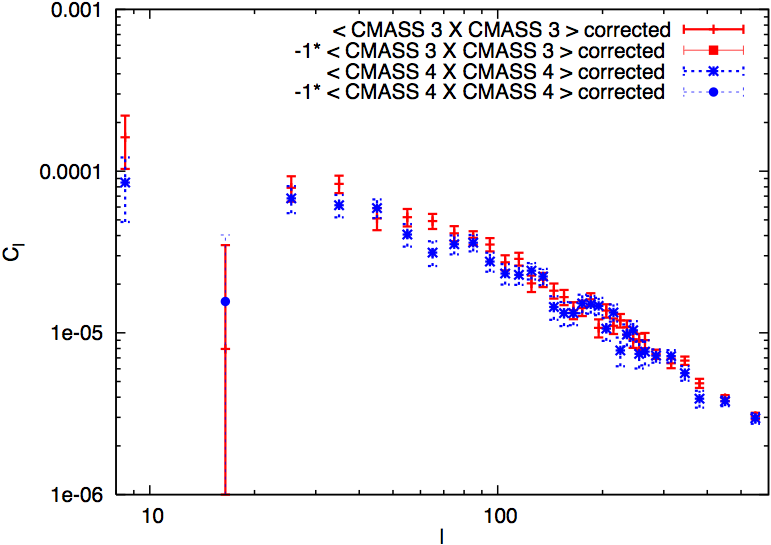}
\end{center}
\caption{ We show the systematic corrected power-spectra of various redshift slices when we include all three dominant systematics (stars, sky and seeing). }
\label{fig:totsyscor}
\end{figure}

Finally, we calculate the final systematic-corrected power by
including all three systematics which are found to have significant
correlations with the observed galaxy density field. 
We include the auto-power (of both systematics and galaxy density fields) and 
all cross-powers (among systematics and galaxy fields) at each angular scale. 
The final corrected power is shown in Figure~\ref{fig:totsyscor}.

Nevertheless, as the optimal quadratic estimator produces optimal
errors and unbiased measurement only when the field is gaussian. 
In the case of highly non-Gaussian fields, the estimate is still
unbiased, but the error is not optimal and may not be accurate.
We can only test the validity of estimated error by the quadratic
estimator if we simulate a large number of systematic mocks, and 
carry out variance tests such as the ones carried out for the galaxy
density fields. 

While we understand the construction of mock galaxy catalogs, we do
not fully understand how to construct  mock systematic fields. 
We lack the 'theory' of systematics fields (except probably the stellar density map). 
Therefore, it is unlikely that we can achieve optimal error on the
systematic corrections within the scope of this analysis. 
The estimated values is unbiased, but the error can be over-estimated
or maybe incorrect \citep{hamilton97}. 
Nonetheless, when the systematic corrections are small, the uncertainty related to the correction cannot be larger than the correction itself.
Therefore, we conclude that the most conservative way would be to include only power from multipoles that have relatively small 
corrections. 
Lacking a better model for the systematics, we adopt the following
simplistic model of estimating the covariance of the 
systematic-corrected power-spectra for multipole bins which require small corrections. 
We assume Gaussianity of the fields involved, and 
thus using the following relationship:
\begin{equation}
\sigma^2(C^{i,j}_l) = \frac{2}{f_{ sky}(2\ell + 1)}(C^i_l + \frac{\Delta\Omega}{N_i}) (C^j_l + \frac{\Delta\Omega}{N_j})
\end{equation}

We modify the above equation by adding the ''correctional power'' due to systematics: 
\begin{equation}
\sigma^2(C^{i,j}_l) = \frac{2}{f_{sky}(2\ell + 1)}\Pi_{k=i,j}(\sqrt{C^2_i(l) + (\Delta{C_i(l)})^2} + N^{shot}_i) 
\end{equation}
This is for each $\delta_{\ell} = 1$, so we take into account the
fact that the $\delta_{\ell}$ is not 1 in all of our bins. 
The quantity $\Delta{C_j(\ell)}$ is the correctional power contribued by systematics.
This method assumes the gaussianity of the fields, which is not a
satisfactory assumption, the optimal quadratic estimator can in principle project out 
powers that are understood, such as those time-dependent systematics
can be projected out in the CMB map-making. 
Nonetheless, a full modeling of the systematics and then projecting them out
using optimal quadratic estimator is a much larger under-taking, which 
will be left to future work.

\begin{figure}
\begin{center}
\leavevmode
\includegraphics[width=3.0in]{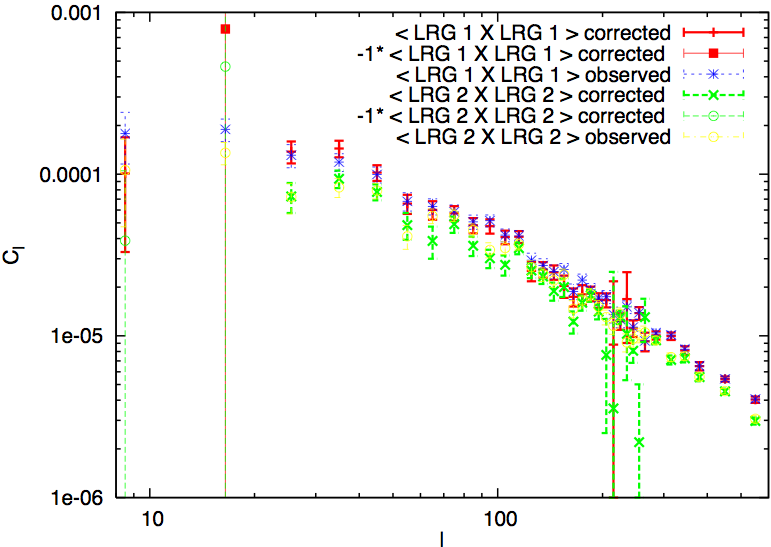}
\includegraphics[width=3.0in]{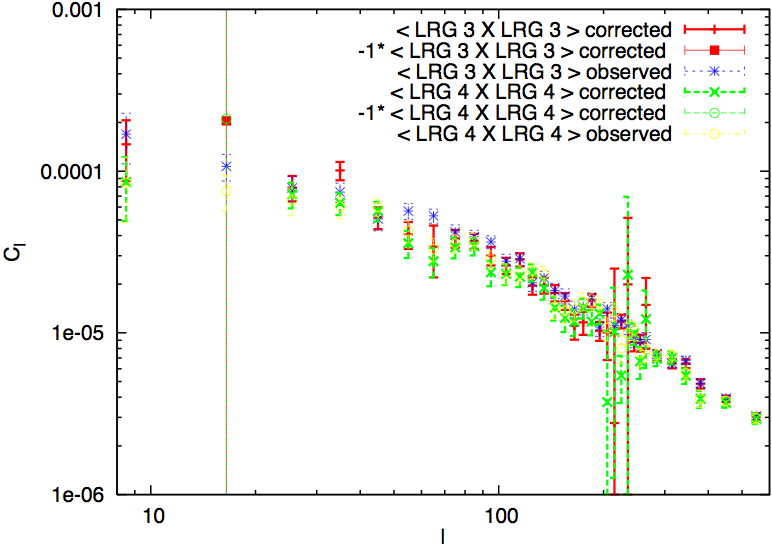}
\end{center}
\caption{ We show the systematic corrected power-spectra and the observed power-spectra (with systematics) of various redshift slices when we include all three dominant systematics (stars, sky and seeing). }
\label{fig:tot_comp_syscor}
\end{figure}

We also show for completeness purposes the power-spectra of various redshift slices before and after the corrections in Figure~\ref{fig:tot_comp_syscor}.

\begin{figure}
\begin{center}
\leavevmode
\includegraphics[width=3.0in]{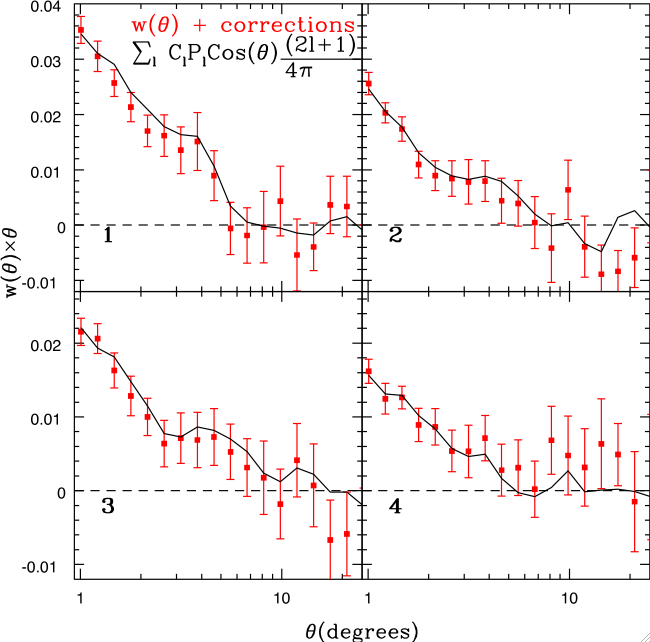}
\end{center}
\caption{ The correlation function computed from
  systematically-corrected angular power-spectra (black lines) is
  compared to the measured and systematically corrected correlation function (red squares with error bars) in Ross et al. (2011). The correlation function in Ross et al. 2011 also are systematically corrected following
our method detailed in this paper. Following the black lines, we can see hints of the BAO at nearly all redshift bins, and that there is 
no significant large scale power seen by other previous analysis such as Thomas et al. (2011)  prior to our DR8 analysis.}
\label{fig:comp_ross}
\end{figure}

We also compute the correlation function of our systematic corrected power-spectra 
and compare with those presented in \cite{ross11} and found that they are completely consistent with each other after systematic corrections.
One of the systematic correlation correction method employed in \cite{ross11}
follows our paper \footnote{Even though this analysis is submitted at a
  later date, the Ross et al. (2011) paper is part of the DR8
  clustering project in SDSS III, and thus Ross et al. (2011) has
  applied our method described here.}, and thus it is not surprising that we have achieved the
same systematically corrected correlation function, even though the computation of the correlation function is independent. The figure~\ref{fig:comp_ross} shows how our computed correlation function from our systematically-corrected optimally-estimated angular power-spectra is 
completely consistent (to within $1.5 \sigma$)  with the measurement
of \cite{ross11}.
We would also note that our correlation function shown in black lines in Figure ~\ref{fig:comp_ross} (our $w(\theta)$ calculated using the angular power-spectra) 
has no significant large scale power, which suggests no exotic
inflation scenario, nor residual systematics. 

\section{Cosmological Parameter Fitting Method and Validation}

\subsection{Method}
\label{sec:cosmo_method}

As described in Sec~\ref{sec:non_linear},  we adopt the simple linear
redshift-independent biasing model (with shot noise subtracted for
every single angular power-spectra). 
Therefore, in addition to the cosmological parameters that are of
interest for each model, we include three extra parameters for each 
redshift slice ($b$, $N_{\rm{shot}}$ and $a$) as shown in \ref{eq:non-linear_eq}.

We have  estimated that the non-linear redshift space distortion
effects are minimal in our case \citep{Saito:2012xx}, 
therefore we include the full linear redshift space distortion following
\cite{padmanabhan07} as discussed in Sec.\ref{sec:theory2}. 
However, calculating the full linear redshift space distortions
requires significant time, and it is different from Limber
approximation at  $l<30$; therefore, we
made a decision to employ Limber approximation for multipole ranges at
$l \gtrsim 30$, and employing the full linear redshift space distortion calculation only at $l<30$. 

The measured band powers from the quadratic estimator have
contributions from a range of wave numbers, even though they are
highly concentrated in their own $\ell$-bin. 
The quadratic estimator is designed to compute nearly anti-correlated
power spectra \citep{padmanabhan03} across different multipole bins,
but it still has a very small ($< \sim 5\%$) contribution from 
other multipole bins.
We take this effect into account by convolving the theory
power-spectra with the window function before calculating the likelihood by $(d-t)^{T} C^{-1} (d-t)$, where $d$ represents the measured power-spectra, $t$ represents the theory power-spectra convolved with window function and $C$ is the covariance across different bands and redshifts as output from the quadratic estimator.

The fitting of the all cosmological parameters are done through MCMC with COSMOMC \citep{cosmomc}.
As it was discussed earlier, we do not know the accuracies of the
error estimated using a highly non-Gaussian field with quadratic estimator. 
We know that it won't be biased, but the error-bar can be
significantly mis-estimated. 
When the systematic corrections are small, we can assume that the error involved with the correction cannot be significant. 
However, at the largest angular scales, some of the systematic
corrections are large enough that only proper error propagation that
involves significant undertaking in full modeling of systematics would provide sufficient accuracy on the error of the correction. 
There are also concerns about the validity of the assumption of linear effects of systematics on the galaxy density field, especially when the corrections are large. 
We do not have indications to believe that the effects of systematics
on galaxy density field is linear or non-linear. 
Therefore, we decided to avoid the scales that involve large systematic corrections, which are concentrated at the large scales, thus setting the start of the multipole
range from $l=30$. 

Since our data-set is a photometric sample, the non-linear effects are mitigated by the 
fact that when we examine the data-set, it is already integrated along
all line-of-sights, thus decreasing its non-linearities. 
We therefore apply the HALOFIT routine for computing the non-linear power-spectrum, and limit ourselves to multipoles that are at relatively large scale, while 
non-linearities remain a small effect. 
We choose the multipole range by simply testing variety of ranges using our simulations described in Sec~\ref{sec:simulations}. 
In order for MCMC chain of the simulation to converge quickly, we 
average the simulations across different simulation boxes (so that they are not correlated). 
We tested a large range of multipole ranges: $1<l<600$, $20<l<150$, $20<l<300$, $20<l<200$ for example, and found $20<l<200$ and $20<l<150$ to return results within 1-sigma of the input 
parameters.
It is also of interest to show the corresponding $k$-limit for the
various $\ell$ ranges, since the range in $k$ may provide an easier
reference for the non-linearities of interest. We list the corresponding $k$ range for the $\ell$ range we use for the redshift bins considered in our analysis in Table~\ref{tab:ktab}.

\begin{table}
\begin{tabular}{cccc}
\hline
Label & $z_{mid}$ & $k_{max}=l_{max}/r(z_{mid})$ &  $k_{max}=l_{max}/r(z_{mid})$ \\
      &           &  $\ell_{max} = 200$  &   $\ell_{max} = 150$ \\
      &           &   $(h^{-1}Mpc)$ &    $(h^{-1}Mpc)$ \\
\hline
\input{kdata.tbl}

\hline
\end{tabular}
\caption{\label{tab:ktab} Descriptions of the 4 $\Delta z=0.05$ redshift slices,
$z_{mid}$ is the midpoint of the redshift interval. We also show the corresponding $k_{max}$ corresponding to the $\ell_{max}$ considered for each redshift slices.}
\end{table}

Combining both the low and the high $l$ limit, we conclude that $30<l<150$ and $30<l<200$ are both  conservative choices for fitting of cosmological parameters.

\subsection{Cosmological Parameter Fitting Method Validation}
In this Section we present some of the tests used to check
$C_l$-likelihood routine for COSMOMC. To perform such a test we need an angular power spectrum whose \textit{input} cosmology (i.e. the cosmological model with which was generated, modulo cosmic variance) is completely known. Therefore, the mock angular power spectrum described in Section~\ref{sec:est_theory}, being derived from a cosmological simulation with initial conditions given by a known set of values for the cosmological parameters, provides an excellent testbed for our fitting routine.

The most straightforward test is to fit each individual angular power spectrum from the mocks and check that every one of them (out of the 160 available) returns the \textit{input} cosmology. 
However, running 160 MCMC chain with only one mock power-spectra each is computationally intensive, especially since each angular power-spectra has the power of $\sim 0.5-1$ actual redshift
slice from the data, thus it will take significant time for the chains to converge if they converge at all.

We therefore need to combine these mock $C_l$ with additional
data-set, and the most obvious choice being the CMB data from WMAP7 \citep{wmap7}. 
The WMAP7 best fit parameters however are not exactly the same as the simulation \textit{input} parameters, thus we replace the standard CMB-likelihood by a much simpler one, in which we compute the value of $\chi^2$ from the actual covariance matrix from WMAP7, but not using the actual parameters themselves. 
To fully validate the fitting method, we need to use mocks that show a similar signal-to-noise ratio as those observed in the data. 
We combine individual mocks by using different simulations (and not different lines of sight in the same simulation).

\subsubsection{Building Covariance of mocks}
We compare the Gaussian covariance matrix of power spectrum from OQE (i.e., `OQE covariance matrix') with the dispersions among 160 mocks (i.e., \Nb covariance matrix').
Note that the 160 mocks are not strictly independent from each other, as different lines of sight share a small but nonzero amount of volume. To exclude the artificial covariance between different lines of sight, we derive the covariance matrix of the 20 independent mocks per each line-of-sight; we then average the eight covariance matrices for eight lines of sight. As a comparison, a straightforward dispersion among 160 mocks gives an almost identical result, implying that the different lines-of-sight share very little volume.

\begin{figure}
\begin{center}
\includegraphics[width=3.0in]{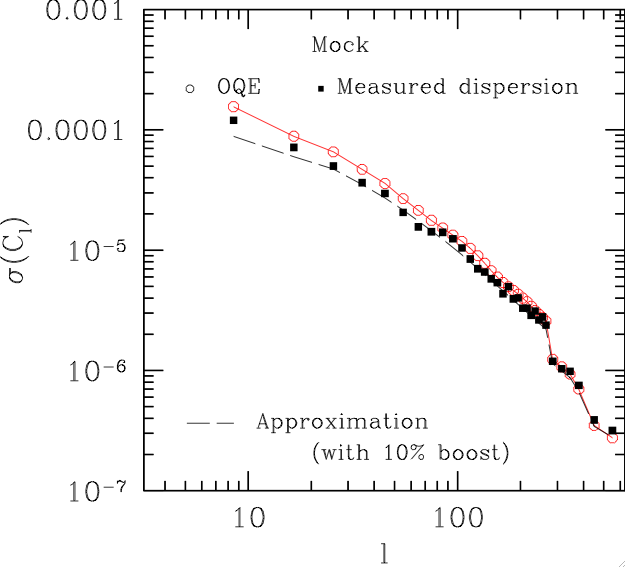}
\includegraphics[width=3.0in]{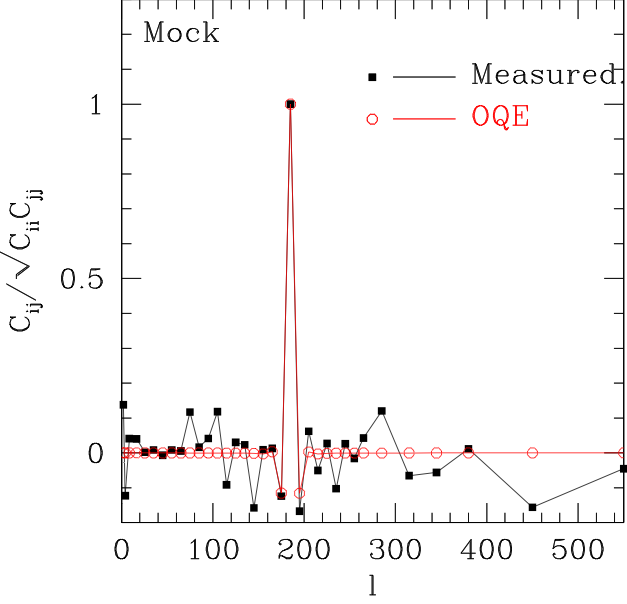}
\end{center}
\caption{Comparisons between the OQE covariance matrix (red circles and solid line) and the dispersions among 160 mocks (black squares).  The black dashed lines are an theoretical, expected error derived based on Gaussianity after rescaled with an empirical boost factor of 1.1 to better match the observed dispersion. The lower panel shows the off-diagonal elements of the \Nb\ covariance matrix in comparison to the OQE covariance, for a slice at $l=185$.  }\label{fig:covmock}
\end{figure}

In the upper panel of Figure \ref{fig:covmock}, the red points are square roots of the diagonal elements of the OQE covariance matrix, and the black squares are from the \Nb\ covariance matrix. The diagonal elements of the OQE covariance matrix can be analytically calculated based on the smooth fit to the measured power spectrum  and the number of independent modes for each wave number band assuming Gaussianity, if the matrix is diagonal. However, the OQE covariance matrix includes the effect of the window function due to the survey geometry, and the covariance matrix therefore is not strictly diagonal and has a small anti-correlation between neighboring bins. Indeed, we find that there is a small deviation between the OQE covariance matrix and a naive Gaussian error calculation without accounting for the window function. The difference is expected since the naive Gaussian error calculation does not include the effects of the actual survey geometry. The black dashed lines in the figure are the theoretical, expected error derived based on Gaussianity; we have rescaled it with an empirical boost factor of 1.1 to better match the observed dispersion. The dispersions between mocks are systematically lower than the OQE on large scales but appears to lie between the OQE expectation and the boosted Gaussian apprixomation.

The lower panel shows the off-diagonal elements of the \Nb\ covariance matrix in comparison to the OQE covariance, for a slice at $l=185$. We observe fluctuations up to 20\% in the measured off-diagonal terms but find no obvious indication that it disagrees with the OQE covariance matrix. Therefore, we conclude that the OQE covariance matrix based on the Gaussian assumption does not underestimate the true error of the 2-dimensional projection of the nonlinear galaxy field.

For the real data, we use the covariance matrix from the OQE for the auto-power for each redshift bin.; while we use the OE The upper panel of Figure \ref{fig:covdata} shows the square roots of the diagonal elements of the OQE covariance matrix of the real data (open circles) in comparison to the prediction based on the Gaussian prediction (after boosted by 1.1: solid squares) for all four redshift bins. The agreement is even better than the mock case, and it is probably due to the larger survey area of the real data that decreases the cross-correlation between different $\ell$ bins that our simple  Gaussian approximation cannot access. The lower panel shows the cross-correlation between different $\ell$ bins for four different slices of the OQE covariance matrix. We overplot covariance matrices for \LRGs\ (black), \LRGe\ (red), \LRGn\ (blue), and \LRGt\ (magenta). Note that the covariance structure is identical for the four redshift bins, which is reasonable as the four redshift bins are subject to the same mask. The same structure should apply to the covariance between different redshift bins as well. We therefore build the cross-covariance between different redshift bins by combining the diagonal elements from the Gaussian assumption and the covariance structure in the right panel of Figure \ref{fig:covdata}.  The diagonal elements are constructed using smooth fits to the measured auto and the cross-power spectra of and between redshift bins and boosted by 1.1 based on the results of the auto power spectra:

\begin{equation}
Cov^G_{ii, jj}(\ell,\ell)=a_{\rm fac}\frac{2}{f_{\rm sky}N_{\rm mode}} C_{ij}(\ell)C_{ij}(\ell),\label{eq:CovG}
\end{equation}
\noindent
where $i$ and $j$ indicates a redshift slice, $f_{\rm sky}$ is the fraction of the sky, $N_{\rm mode}$ is the number of wave modes within the band, and $C_{ij}(\ell)$ is a smooth fit to the auto or cross power spectrum; we include the shot noise contribution to $C_{ij}(\ell)$ in the case of the auto power spectra (i.e., $i=j$). The factor $a_{\rm fac}$ is the empirical factor of 1.1 that we introduce to match the OQE covariance matrix and equation \ref{eq:CovG}. We use this equation to build the covariance between different redshift slices, while using the OQE covariance matrix for the covariance within the redshift slice.

\begin{figure}
\label{fig:covdata}
\begin{center}
\includegraphics[width=3.0in]{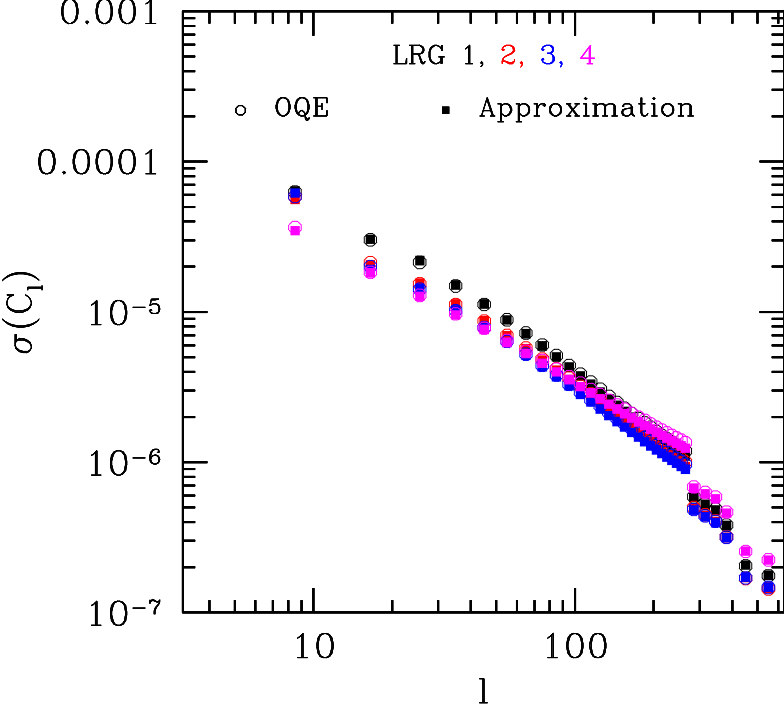}
\includegraphics[width=3.0in]{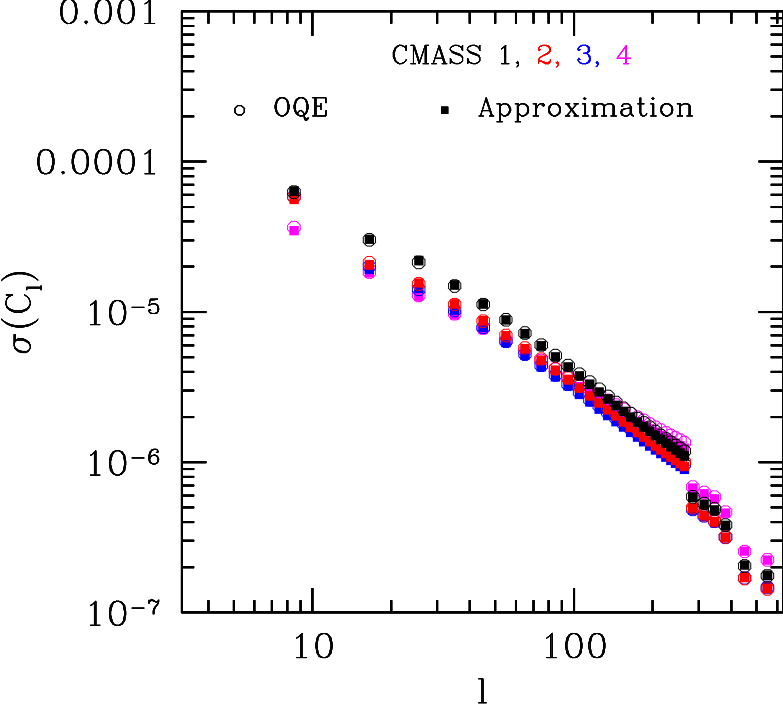}
\end{center}
\caption{The covariance matrix structure for the real data. The top panel compares the OQE prediction (open symbols) and the boosted Wick's theorem (solid symbols) for the four redshift bins. The two agree very well. The bottom panel shows the uniformity of the off-diagonal structure of the four redshift bins for four slices of the covariance matrix. The uniformity arises from the common mask. We therefore use Gaussian assumption and such uniform off-diagonal structure to build a cross-covariance between different redshift bins. }\label{fig:covdata}
\end{figure}

\subsubsection{Mock test results}
With the combined average of 20 spectra each in combination with the
pseudo-WMAP7, we find that the above model
recovers all input cosmological parameters of the CMASS mocks for all 8 averaged power-spectra to within $1.5 \sigma$.
The recovered bias parameters are also very similar to the input bias of the CMASS mocks as described in \cite{white11}.
We therefore conclude that this model accurately recovers cosmological parameters when used in the range of
angular scale specified above.

\section{Results}
\label{sec:results}

\subsection{Constraints on Cosmological Models}

The angular clustering measurement can be used to constrain cosmological model
in several different ways: through standard rulers such as the matter-radiation
turn-over scale, the baryon acoustic oscillations, or through 
large scale power which would constrain primordial non-gaussianities \citep{dalal08,slosar08}. 
In companion paper \citep{seo11}, we examine only baryon acoustic oscillations,
and remove any contribution from the overall shape of the
power-spectrum; 
in this paper, we include the overall shape of the power-spectrum and 
parts of the baryon acoustic oscillations to derive constraints
on cosmological models. There is a companion publication 
on the neutrino mass constraints using the same angular power-spectrum \citep{deputter11}.

Here,  we choose to consider a variety of cosmological models, although not intending to exhaust all possibilities. 
We include several other data-sets to help break cosmological degeneracies, such as WMAP7 \citep{wmap7}, 
the ``Union 2'' supernova dataset (hereafter SN), which includes 557 supernova from \cite{hamuy96,riess99,riess07,astier06,jha06,woodvasey07,holtzman08,hicken09,kessler09}, and  $H_0$ constraints from using 600 Cepheids observed 
by Wide Feild Camera 3 (WFC3) published by \cite{riess11} (HST).

\subsubsection{Flat CDM model with a constant equation of state} 

\begin{figure}
\begin{center}
\leavevmode
\includegraphics[width=3.0in]{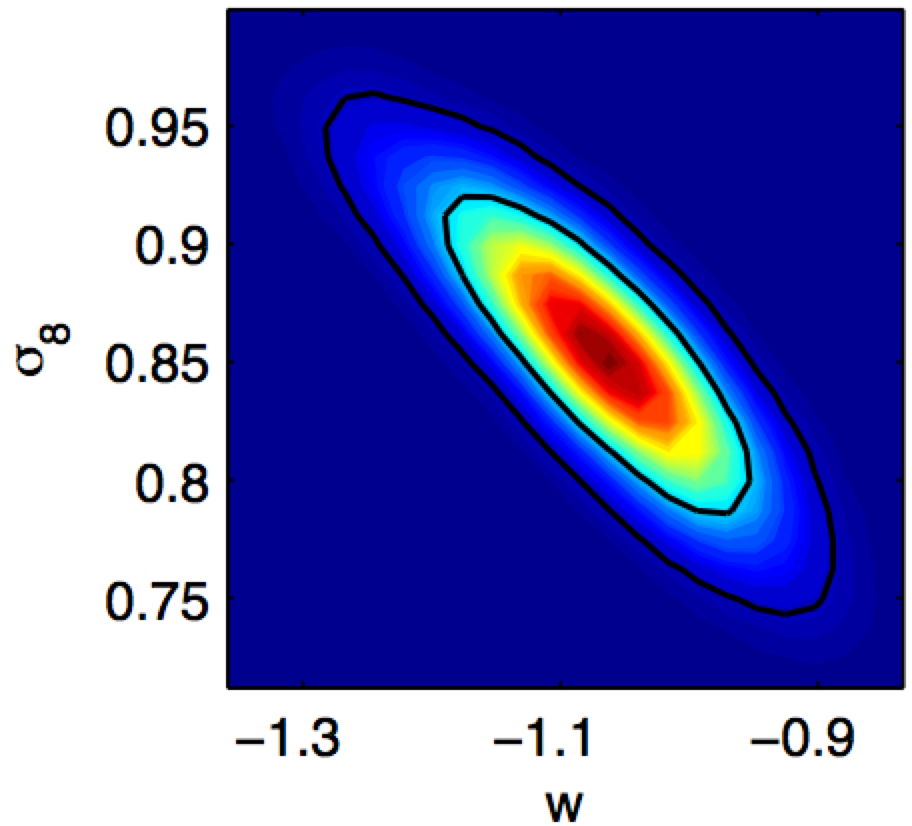}
\end{center}
\caption{ The 2D contour of $\sigma_8$ and $w$ when combining WMAP7 +
  HST + DR8. The results are consistent with $w=-1$.}
\label{fig:wcdm_2D}
\end{figure}

We investigate the flat CDM model with a constant equation of state parameter ($w$) to 
characterize Dark Energy with the combination of our angular
power-spectra from SDSS-III Data Release 8 (DR8) with 
other data-sets.
When we combine our DR8 observed angular power-spectra with WMAP7 +HST
+ SN dataset, we find $w=-1.07\pm0.0775$, $\Omega_m=0.2699\pm 0.0166$
and $\sigma_8 = 0.85 \pm 0.044$ (see Fig~\ref{fig:wcdm_2D}).
We also combine our systematic-corrected DR8 angular power-spectra
with WMAP7 +HST +SN data-set, we find $w=-1.064\pm0.0757$,
$\Omega_m=0.267\pm0.0163$. The systematically-corrected
angular power-spectra gives consistent results compared with the observed angular
power-spectra. 


We compare our results with other large scale structure datasets, 
such as the latest large scale structure constraints from galaxy clustering in \cite{blake10}, which 
has detected BAO at $z\sim 0.6$ using spectroscopic survey WiggleZ which includes 200,000 galaxy spectra over 800 $\rm{deg}^2$.
They found a similar constraints on the equation of state
of dark energy: $w=-1.03\pm 0.08$ when they combined with WMAP7 + SN.
This implies that our dataset, even though it is purely imaging data,
gives a similar constraining power when compared to latest spectroscopic
survey such as WiggleZ.

We also compare our results with the BAO constraints from SDSS-DR7, when combined
with WMAP7 + HST \citep{reid10,percival10}, they found $w=-1.10\pm
0.14$, while \cite{montesano11} used full shape of P(k) from SDSS-DR7,
when they combined it with CMB + HST,  they found $w = -1.07 \pm 0.11$. 
When we combine 
with the same dataset (CMB+HST), we find $w=-1.165\pm 0.12$, which
implies that our dataset gives a similar constraining power as with the full 3D DR7 spectroscopic
sample (at $z<0.45$), while our purely imaging dataset is at a higher redshift range
(0.45-0.65). 

\subsubsection{Open $\Lambda$CDM model} 

\begin{figure}
\begin{center}
\leavevmode
\includegraphics[width=3.0in]{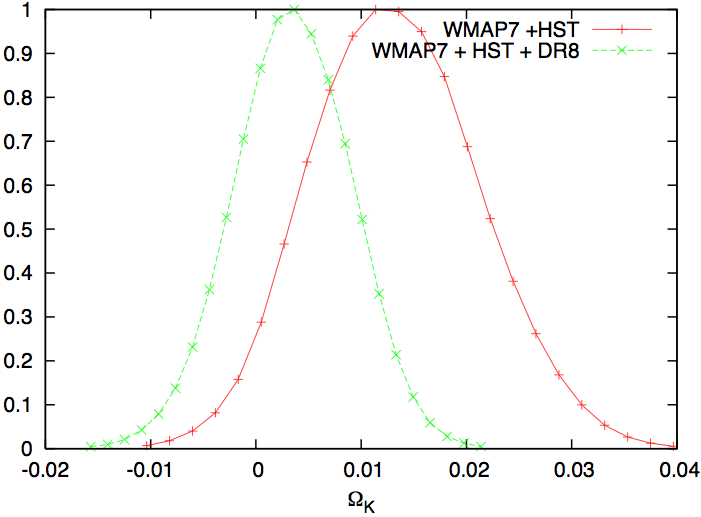}
\end{center}
\caption{ A comparison of the constraints on the flatness of the Universe using our angular power-spectrum from SDSS-DR8 (+WMAP7 + HST) versus
WMAP7+HST alone for an open $\Lambda$CDM model.}
\label{fig:ocdm_1D}
\end{figure}

For an open CDM model, when combined with WMAP7+HST, we find $\Omega_K
= 0.00348\pm0.00539$, improving the accuracy over the WMAP7 +HST constraints on 
$\Omega_K$ by $40\%$ ($\sigma(\Omega_K) = 0.0076$ from WMAP7+HST);  see Fig~\ref{fig:ocdm_1D}. 

When we compare our measurement to other large scale structure
measurements such as DR7-BAO constraints; 
we find very similar constraints; for example, on $\Omega_K$ as WMAP7+HST+DR7 gives $\Omega_K=-0.0023\pm0.0055$.

\subsubsection{Flat $\Lambda$CDM model}
\begin{figure}
\begin{center}
\leavevmode
\includegraphics[width=3.0in]{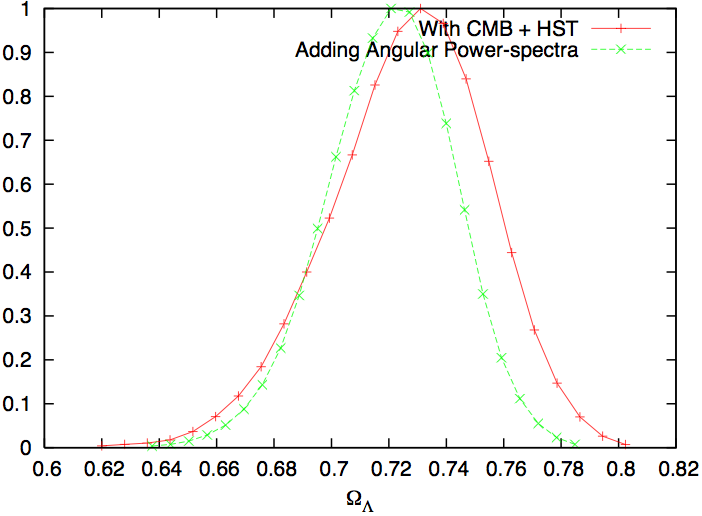}
\includegraphics[width=3.0in]{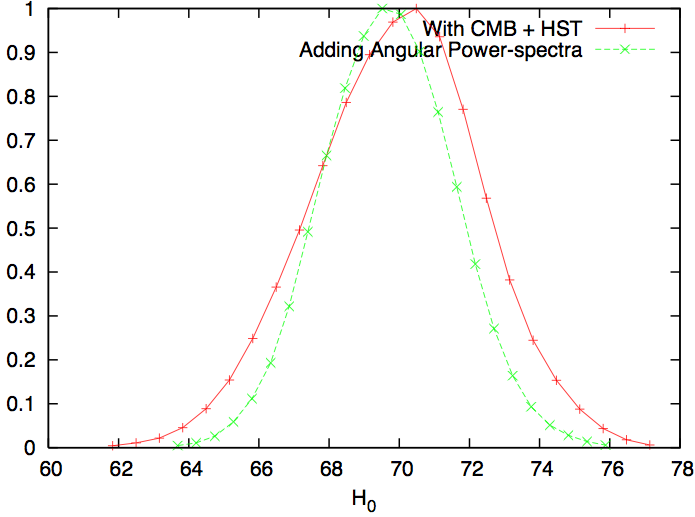}
\end{center}
\caption{ The 1D marginalized constraints of $\Omega_\Lambda$ and $H_0$ when compared to using only WMAP7 + HST. }
\label{fig:lcdm_1D}
\end{figure}

For a flat $\Lambda$CDM model, when combined with WMAP + HST, we find $\Omega_\Lambda
 = 0.73 \pm 0.019$, $\sigma_8$ to be $0.817 \pm 0.023$ and $H_0$ to be
 $70.5\pm1.6$ $ \rm{s}^{-1} \rm{Mpc}^{-1}\rm{km}$, which are
 consistent with WMAP+HST only, while improving the accuracy over
just WMAP+HST by $\sim 5\%$ for all parameters. 
We show the improvement on cosmological constraining power over the combination of WMAP7 and HST in Figure~\ref{fig:lcdm_1D}.

\subsection{Companion Results}
\begin{figure}
\begin{center}
\includegraphics[width=3.0in]{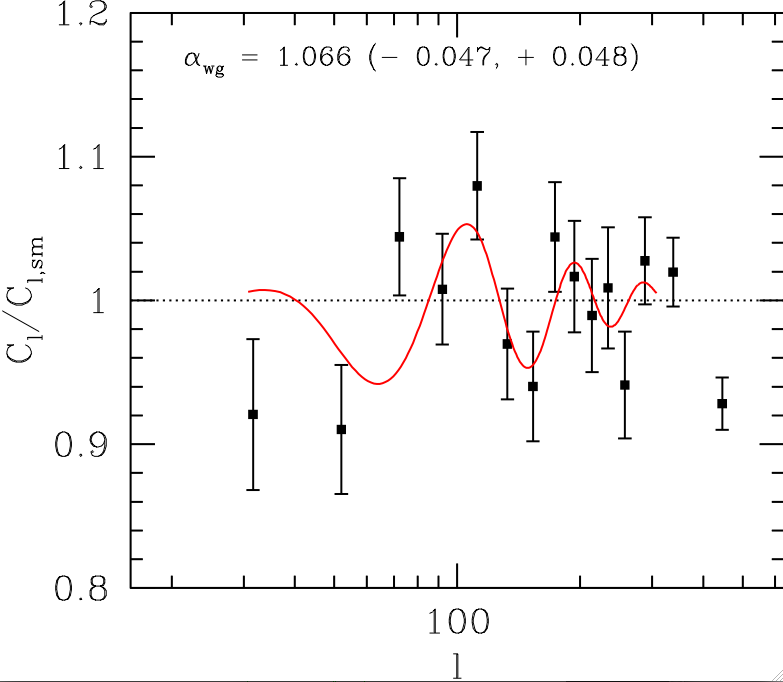}
\includegraphics[width=3.0in]{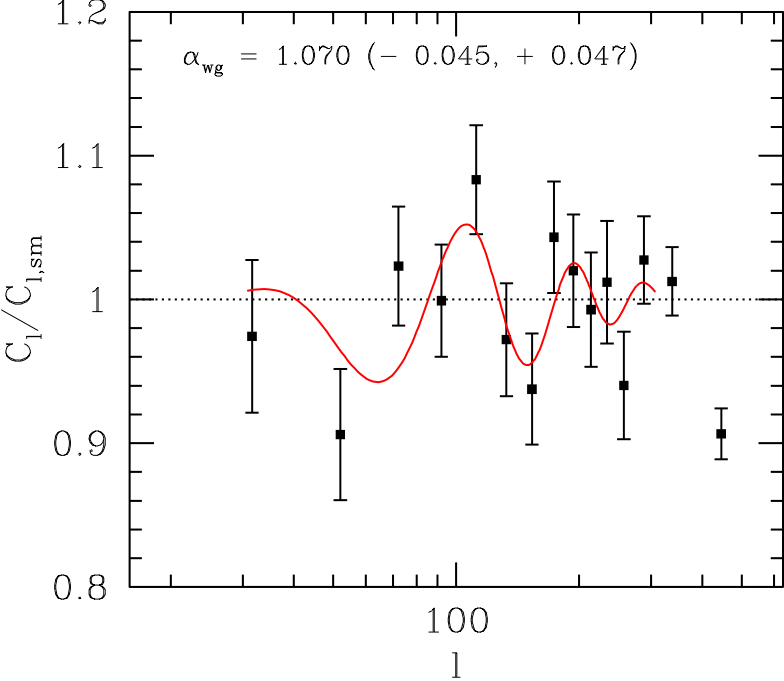}
\end{center}
\caption{The BAO-only fits derived in Paper II: a stacked $C_l/C_{\rm l,sm}$ data of the four redshift bins before (left) and after systematics correction(right). $\alpha_{\rm wg}$ means the best fit $\DArs/8.584 {\rm Mpc}$ each case. The solid red line is the best fit for LRG2, as a comparison, after the wavenumber is rescaled to $z=0.54$. For more details, see Paper II.}\label{fig:BAOonly}
\end{figure}

In this paper, we are utilizing the full power spectum over $30<l<150$, including both the broadband shape and the BAO. As a comparison, our companion paper Paper II, Seo et al. (2011) derives the angular diameter distance scale using the BAO feature over $30<l<300$ as a standard ruler, while excluding nearly all the non-BAO information. To summarize their method and result, they use the angular power spectra and the covariance matrix shown in this paper, build a reasonable template power spectra based on the estimated, true galaxy distribution as a function of redshift and the concordance cosmology, and fit for the distance scale, while marginalizing over many free parameters that account for the shape of the broad band. They derive $\DArs= \afit{9.212}{0.416}{0.404}$ at $z=0.54$ and the result is shown to be robust against assumptions they make during the fitting process. Figure \ref{fig:BAOonly} summarizes the BAO fits they derived before and after the systematics correction.

In a second companion paper, \cite{deputter11}, the angular spectra discussed in the present paper are are used to derive a strong new upper bound on the sum of neutrino masses. As neutrinos suppress growth of structure on scales above the neutrino free streaming length, they leave a characteristic signature in the power spectra. To exploit this signal, \cite{deputter11} model the galaxy spectra and their neutrino mass dependence, test the model using mocks and show that it can be safely applied to the multipole range $\ell = 30-200$, while also considering the conservative range $\ell = 30-150$. The angular clustering galaxy data are then combined with priors from WMAP7, the HST Hubble parameter measurement, supernova distances and the (low redshift) SDSS BAO measurement and the resulting upper bounds are discussed. We quote here the conservative bound, from DR8+CMB+HST+BAO+SN, of $\Sigma m_{\nu} < 0.35$eV at $95 \%$ CL, and refer the reader to the article itself for more details.

\section{Conclusion and Discussion}
\label{sec:discuss}
We have measured the 2D clustering power spectrum of luminous 
red galaxies using the SDSS-DR8 photometric survey. 
The principal results of this analysis are summarized and discussed below.

Using photometric redshifts, we constructed a large uniform sample of galaxies between 
redshifts $z=0.45$ to $0.65$. This probes a cosmological volume of $\sim 3 
h^{-3} {\rm Gpc}^{3}$, making this the largest cosmological volume 
ever used for a galaxy clustering measurement.  The large volume allows us
a measurement of power going from the smallest scales to the largest. In particular, we probe all the way from the smaller scales 
such as the Baryon Acoustic Oscillations scale out to the scale of matter-radiation equality with 
one of the most accurate measurement of angular clustering at $z=0.45-0.65$ achieved to date. 

We applied a novel approach of treating systematics by incorporating both the cross/auto-power of systematics with themselves and cross-power with 
galaxies. This allows us to not only understand the impact of various systematics on observed galaxy number densities, but also 
allows applcation of small corrections at scales where the corrections are small, thus the uncertainty related to the corrections are negligible. 
Since we choose only scales that are minimally affected by systematics, we expect the final cosmological constraints 
from both pre-/and post-systematic corrections are consistent with each other, which is indeed the case. 
This method can be improved drastically by two extra components, which will be left to future work. 
First, we should be able to project out the appropriate modees that are contributed by the systematics in similar way as it is done in CMB map-making \citep{stompor02}, and this can be done when one is estimating the optimally estimated power-spectra. 
Second, we should be able to model the distribution of the systematics (for example, whether it is gaussian or not) by investigating multi-epoch 
data that is available in SDSS-DR8. In particular, this is available in the Stripe 82 area, which even though small (in comparison to DR8 full foot-print), it contains $\sim 250 $ square degrees which were multiply scanned ($\sim 15-20$ repeats for each fields). This will allow us to estimate the undertainty 
of our systematic corrections properly. These two components will significantly improve not only our understanding of the systematics, thus allowing us to 
push our analysis to include larger angular scales (which are affected by the systematics more significantly). However, both of them requires significant
undertakings in both data collection, code development and simulations, while the improvements will not dramatically change our cosmological interpretation
in this paper, therefore, we leave these two components for future projects.  

For a flat $\Lambda$CDM model, combining our data with WMAP7 + HST, we
find $\Omega_\lambda = 0.7301 \pm 0.019$ and $H_0$ to be $70.5\pm1.6$
$ \rm{s}^{-1}\rm{Mpc}^{-1}\rm{km}$.
For a open $\Lambda$CDM model, when combined with WMAP7 + HST, we find $\Omega_K =  0.003476 \pm 0.00538$, improved over WMAP7+HST alone by $40\%$.
For a wCDM model, when combined with WMAP7+HST+SN, we find $w = -1.071 \pm  0.0775$, and $H_0$ to be $71.31 \pm 1.65$ $ \rm{s}^{-1}\rm{Mpc}^{-1}\rm{km}$, which is competitive with the latest large scale structure constraints from spectroscopic surveys such as those by WiggleZ \citep{blake10}, and SDSS DR7 spectroscopic surveys, especially in analysis led by \citep{reid10,percival10,montesano11}.
This results implies that our dataset, even though it is purely
imaging data, possesses a similar constraining power as the spectroscopic surveys such as WiggleZ or SDSS-DR7. What we lack in redshift precision, we compensate by shear volume.  This suggests that future and upcoming imaging surveys such as
PanStarrs \footnote{http://pan-starrs.ifa.hawaii.edu/public/},  DES
\footnote{http://www.darkenergysurvey.org/science/index.shtml} and  LSST \footnote{http://www.lsst.org} can achieve significant cosmological constraints via large scale structure clustering even when compared to other spectroscopic surveys.

This is the Paper I of the project, which mostly describes the construction of the data-set, treatment of systematics, estimation of the angular 
power-spectra and finally using the overall shape of the angular power-spectra over a large range of angular scale to derive 
constraints on our cosmological models. We refer readers to Paper II \citep{seo11}  of the project, which uses only the Baryon Acoustic Oscillation feature 
to fit for various cosmological parameters. We also refer to Paper III \citep{deputter11} of the project, which uses the overall shape of the 
power-spectra to fit for various neutrino models.


%


\section{Acknowledgment}

We would like to thank Nico Hamaus for testing our treatment for non-linearities using N-body simulations, 
Pat McDonald for fruitful discussion on our systematic treatments and Uros Seljak for useful discussions.
S.H. would like to acknowledge Martin White for all the useful
discussions and encouragement even though he is already a co-author.
S.H. would also like to thank UC Berkeley, Department of Energy and Lawrence Berkeley National Laboratory for ther support through Seaborg and Chamberlain Fellowship.
This work was supported by the U.S. Department of Energy under
Contract No. DE-AC03-76SF00098 and in part by the facilities and staff of the Yale University Faculty of Arts and Sciences High Performance Computing Center.
This research used resources of the National Energy Research Scientific Computing Center, which is supported by the Office of Science of the U.S. Department of Energy under Contract No. DE-AC02-05CH11231.
Funding for SDSS-III has been provided by the Alfred P. Sloan Foundation, the Participating Institutions, the National Science Foundation, and the U.S. Department of Energy Office of Science. The SDSS-III web site is http://www.sdss3.org/.
         
SDSS-III is managed by the Astrophysical Research Consortium for the Participating Institutions of the SDSS-III Collaboration including the University of Arizona, the Brazilian Participation Group, Brookhaven National Laboratory, University of Cambridge, Carnegie Mellon University, University of Florida, the French Participation Group, the German Participation Group, the Instituto de Astrofisica de Canarias, the Michigan State/Notre Dame/JINA Participation Group, Johns Hopkins University, Lawrence Berkeley National Laboratory, Max Planck Institute for Astrophysics, New Mexico State University, New York University, Ohio State University, Pennsylvania State University, University of Portsmouth, Princeton University, the Spanish Participation Group, University of Tokyo, University of Utah, Vanderbilt University, University of Virginia, University of Washington, and Yale University.

\bibliography{lrgdr8}
\bibliographystyle{mnras}

\appendix

\section{Quadratic Estimator}

\label{sec:quad_imple}

Consider a Gaussian random field $x_i$ with $\langle x_i\rangle=0$
and covariance
\begin{equation}
 \left\langle x_ix_j\right\rangle = C^{(0)}_{ij} +
    \sum_{\alpha=1}^N p_\alpha C^{(\alpha)}_{ij}
\end{equation}
We wish to form an estimator, $\widehat{p}_\alpha$, of $p_\alpha$ which
is quadratic in the data
\begin{equation}
 \widehat{p}_\alpha = \sum_{ij} Q^{(\alpha)}_{ij} x_i x_j - b_\alpha
\end{equation}
where $Q$ is symmetric.
Requiring the estimator to be unbiased
\begin{equation}
 \left\langle\widehat{p}_\alpha\right\rangle = p_\alpha
 \quad \Rightarrow \quad
 {\rm tr}\left[ Q^{(\alpha)}C^{(\beta)} \right]=\delta^{\alpha\beta}
 \quad , \quad
 b_\alpha = {\rm tr}\left[ Q^{(\alpha)}C^{(0)} \right]
\end{equation}
For Gaussian $x_i$ the covariance of $\widehat{p}_\alpha$ is
\begin{equation}
 {\rm Cov}\left[\widehat{p}_\alpha,\widehat{p}_\beta\right] =
 \sum_{ijkl} Q^{(\alpha)}_{ij} Q^{(\beta)}_{kl}
 \left[ C_{ik}C_{jl} + C_{il}C_{jk} \right]
 = 2 {\rm tr}\left[ CQ^{(\alpha)}CQ^{(\beta)}\right]
\end{equation}
This problem is easiest if we consider a single parameter at a time, with
all other parameters held fixed (and absorbed into $C^{(0)}$).
Thus we wish to minimize
\begin{equation}
 {\rm tr}\left[ CQ^{(\alpha)}CQ^{(\alpha)} - 2\lambda C^{(\alpha)}Q^{(\alpha)}
   \right] \quad .
\end{equation}
Taking derivatives with respect to the components of $Q^{(\alpha)}$ gives
\begin{equation}
 CQ^{(\alpha)}C = \lambda C^{(\alpha)}
\end{equation}
or
\begin{equation}
 Q=  (2F)^{-1} C^{-1} C^{(\alpha)} C^{-1}
  =  \left({\rm tr}
     \left[C^{-1}C^{(\alpha)}C^{-1}C^{(\alpha)}\right]\right)^{-1}
     C^{-1} C^{(\alpha)} C^{-1}
\end{equation}
If the dependence of $C$ on $p_\alpha$ is not linear then we can use a
Newton-Raphson iteration where now $C^{(\alpha)}$ is the derivative of $C$
evaluated at the current best value of $p$.
Iterating, by replacing $p\to p+\delta p$ until the best-fit $\delta p=0$,
results in a maximum likelihood solution.
In practice, it only takes a few iterations to achieve the maximum
likelihood solution. 

This approach also results in another fact that is under-appreciated in the
literature.
The above choice of $Q$ (which can have a slightly different form, see
Table 4 of \cite{padmanabhan03} for more details) produce
error-bars that are anti-correlated across different band powers. 
In this paper, we include the window function (which is mostly
affected by the mask) before we compare the observed power and the
theoretical power.  


%
%
%


\end{document}

%% file: lrgdata.tbl
CMASS 0 &0.425  &   23517&     \\
CMASS 1 &0.475  &  154531&    1.92$\pm$  0.06 \\
CMASS 2 &0.525  &  198132&    1.98$\pm$  0.05 \\
CMASS 3 &0.575  &  190603&    2.19$\pm$  0.05 \\
CMASS 4 &0.625  &  121181&    2.19$\pm$  0.05 \\
CMASS 5 &0.675  &   54744&    \\

%% file: kdata.tbl
LRG 1 &0.475  &0.160  & 0.119    \\
LRG 2 &0.525  &0.145  & 0.110    \\
LRG 3 &0.575  &0.134  & 0.100    \\
LRG 4 &0.625  &0.125  & 0.094    \\